\documentclass[12pt]{article}
\usepackage{enumerate}
\usepackage{amsfonts}
\usepackage{amsmath}
\usepackage{amssymb}
\usepackage{amsthm}
\usepackage{latexsym}
\usepackage{color}

\def\H{\mathcal{H}}

\def\K{\mathcal{K}}

\def\S{\mathfrak{S}}

\def\N{\mathbb{N}}
\def\Ch{\mathrm{Ch}}
\def\F{\mathfrak{F}}
\def\T{\mathfrak{T}}
\def\B{\mathfrak{B}}

\newcommand{\supp}{\mathrm{supp}}
\newcommand{\rank}{\mathrm{rank}}
\newcommand{\id}{\mathrm{Id}}
\newcommand{\Tr}{\mathrm{Tr}}

\newcommand{\shs}{\hspace{1pt}}
\newcounter{defin}  \newcounter{lemma}  \newcounter{theorem}
\newcounter{property} \newcounter{corol}  \newcounter{remark} \newcounter{example}
\newenvironment{lemma}{\par\refstepcounter{lemma}     \textbf{Lemma \thelemma.} }{\rm\par}
\newenvironment{theorem}{\par\refstepcounter{theorem}     \textbf{Theorem \thetheorem.}\ }{\rm\par}
\newenvironment{property}{\par\refstepcounter{property}     \textbf{Proposition \theproperty.}\ }{\rm\par}
\newenvironment{corollary}{\par\refstepcounter{corol}     \textbf{Corollary \thecorol.} }{\rm\par}

\newenvironment{remark}{\par\refstepcounter{remark}     \textbf{Remark \theremark.}}{\rm\par}
\newenvironment{example}{\par\refstepcounter{example}     \textbf{Example \theexample.}}{\rm\par}

\begin{document}

\title{Compactness criterion for families of quantum operations in the strong convergence topology and its applications}

\author{M.E. Shirokov\footnote{email:msh@mi.ras.ru}\\Steklov Mathematical Institute, Moscow}
\date{}
\maketitle



\begin{abstract} A revised version of the compactness criterion for families of quantum operations in the strong convergence topology
obtained in \cite{AQC} is presented, along with a more detailed proof and the examples showing the necessity of this revision.

Several criteria for the  existence of limit points of a sequence of quantum operations w.r.t. the strong convergence are obtained and discussed.

Applications in different areas of quantum information theory are described.
\end{abstract}


\tableofcontents

\section{Introduction and preliminaries}

In the study  of finite-dimensional quantum channels and operations  the diamond norm distance between them
is widely used \cite{Kit},\cite[Section 9]{Wilde}. The convergence of quantum channels and operations induced by this distance is naturally called the \emph{uniform convergence}.

In the analysis of infinite-dimensional quantum channels and operations the diamond norm distance and the uniform convergence are also used (see, f.i.\cite{H&W,Pir,Wilde-sc}), but in general the uniform convergence is too strong and do not reflect
the physical nature of such channels and operations (the most striking example confirming this can be found in \cite{W-EBN}).  In the infinite-dimensional  case it is natural to use the \emph{strong convergence}
 which is the convergence in the strong operator topology on the space of bounded linear maps between Banach spaces of trace-class operators. A sequence of quantum channels (operations) $\Phi_n$ from a system $A$ to a system $B$ strongly converges to a quantum channel (operation) $\Phi_0$ if
\begin{equation*}
\lim_{n\rightarrow+\infty}\Phi_n(\rho)=\Phi_0(\rho)
\end{equation*}
for all states $\rho$ of the system $A$, where $"\lim"$ denotes the limit w.r.t. the trace norm on the set of trace-class operators on the space $\H_B$ describing the system $B$.\footnote{There is a more weak topology on the set of infinite-dimensional quantum operations called the weak$^*$ operator topology \cite[Section III]{L&Co}. Corollary 11 in \cite{L&Co} claims the compactness of  important classes of operations in this topology (which are not compact in the strong convergence topology).}

It seems that the first systematic study of the strong convergence  of  quantum channels and operations was carried out in \cite{AQC},\footnote{I would be grateful for any comments concerning this point.} where this type of convergence was used to develop a method for investigating the information characteristics of infinite-dimensional quantum channels based on approximation. In particular, it is shown in \cite{AQC} that \emph{it is the strong convergence topology that makes the set of all quantum channels (resp. operations) between quantum systems $A$ and $B$ topologically isomorphic to a certain subset of states (resp. positive trace class operators) on the space $\H_{BR}$}, where $R$ is a reference system equivalent to $A$
(the generalized Choi-Jamiolkowski isomorphism). Using this topological isomorphism a simple compactness criterion for families of quantum channels and operation in the strong convergence topology is established and analysed in \cite{AQC}. This compactness criterion has proved useful for solving various problems related to the study of quantum systems and channels of infinite dimension.
Several results obtained in \cite{HCD,CMI,REC} by applying this criterion are described in Section 3 along with its new applications.

Unfortunately, there is an inaccuracy  in the formulation
of Corollary 2A in  \cite{AQC}, where the nontrivial part of the compactness criterion mentioned before is presented. Formally, this  inaccuracy
consists in missing the word "closed" in the first line of that corollary. Fortunately, \emph{it did not affect all the applications} of this
compactness criterion  (known to me and described in Section 3) because in all these applications the compactness criterion was used to prove the \emph{relative} compactness of a certain sequence of quantum channels (operations) w.r.t.  the strong convergence with the aim to show the existence of a limit point of this sequence.

The aim of this article is to give a correct formulation of the compactness criterion for families of quantum channels and operations w.r.t. the strong convergence
with a more detail proof and to describe its different versions and applications.\medskip

Let $\mathcal{H}$ be a separable Hilbert space,
$\mathfrak{B}(\mathcal{H})$ the algebra of all bounded operators on $\mathcal{H}$ with the operator norm $\|\cdot\|$ and $\mathfrak{T}( \mathcal{H})$ the
Banach space of all trace-class
operators on $\mathcal{H}$  with the trace norm $\|\!\cdot\!\|_1$. Let
$\mathfrak{S}(\mathcal{H})$ be  the set of quantum states (positive operators
in $\mathfrak{T}(\mathcal{H})$ with unit trace) \cite{H-SCI,Wilde,BSimon}.

Denote  the unit operator on a Hilbert space
$\mathcal{H}$ by $I_{\mathcal{H}}$ and  the identity
transformation of the Banach space $\mathfrak{T}(\mathcal{H})$ by $\id_{\mathcal{\H}}$.\smallskip

A \emph{quantum operation} $\Phi$ from a system $A$ to a system
$B$ is a completely positive trace-non-increasing linear map from
$\mathfrak{T}(\mathcal{H}_A)$ into $\mathfrak{T}(\mathcal{H}_B)$.
A trace preserving quantum operation is called  \emph{quantum channel} \cite{H-SCI,Wilde}.   \smallskip
For any  quantum operation  $\,\Phi:A\rightarrow B\,$ the Stinespring theorem implies the existence of a Hilbert space
$\mathcal{H}_E$ and  a contraction
$V_{\Phi}:\mathcal{H}_A\rightarrow\mathcal{H}_B\otimes\mathcal{H}_E$ such
that
\begin{equation*}
\Phi(\rho)=\mathrm{Tr}_{E}V_{\Phi}\rho V_{\Phi}^{*},\quad
\rho\in\mathfrak{T}(\mathcal{H}_A).
\end{equation*}
If $\Phi$ is a channel then $V_{\Phi}$ is an isometry. The minimal dimension of $\H_E$ is called the \emph{Choi rank} of $\Phi$  \cite{H-SCI,Wilde}. \smallskip

The quantum operation
\begin{equation}\label{comp-ch}
\widehat{\Phi}(\rho)=\Tr_B V_{\Phi}\rho V^*_{\Phi},\quad \rho\in\T(\H_A),
\end{equation}
from $A$ to $E$ is called \emph{complementary} to the operation $\Phi$ \cite{H-SCI,H-c-ch}. A complementary operation to an operation $\Phi$ is uniquely defined up to the isometrical equivalence \cite{H-c-ch}:  if $
\widehat{\Phi }^{\prime }:\mathfrak{T}(\mathcal{H}_{A})\rightarrow \mathfrak{T}(\mathcal{H}_{E^{\prime }})$ is a quantum operation  defined by (\ref{comp-ch})
via another contraction $V^{\prime }:\mathcal{H}_{A}\rightarrow \mathcal{H}
_{B}\otimes \mathcal{H}_{E^{\prime }}$ then there is a partial isometry $W:
\mathcal{H}_{E}\rightarrow \mathcal{H}_{E^{\prime }}$ such that
\begin{equation*}
\widehat{\Phi }^{\prime }(\rho)=W\widehat{\Phi }(\rho)W^{\ast },\quad
\widehat{\Phi }(\rho)=W^{\ast }\widehat{\Phi }^{\prime }(\rho)W,\quad \rho
\in \mathfrak{T}(\mathcal{H}_{A}).
\end{equation*}

The \emph{strong convergence topology} on the set $\F_{\geq1}(A,B)$ of  quantum operations from $A$ to $B$ is defined by the family of seminorms $\Phi\mapsto\|\Phi(\rho)\|_1$, $\rho\in\S(\H_A)$ \cite{AQC}. The convergence of a sequence $\{\Phi_n\}$ of quantum operations (resp. channels) to a quantum operation (resp. channel)  $\Phi_0$ in this topology means that
\begin{equation}\label{star+}
\lim_{n\rightarrow\infty}\Phi_n(\rho)=\Phi_0(\rho)\quad\forall\rho\in\S(\H_A).
\end{equation}
The strong convergence topology on the set $\F_{\geq1}(A,B)$ is \emph{metrizable}, since it can be defined by the countable family of seminorms $\Phi\mapsto\|\Phi(\rho)\|_1$, $\rho\in\S_0$, where $\S_0$ is any countable dense subset of $\S(\H_A)$. The set $\F_{=1}(A,B)$ of  quantum channels from $A$ to $B$
equipped with the strong convergence topology is a closed subset of $\F_{\geq1}(A,B)$.

An equivalent definition of the strong convergence of quantum channels  is given by Wilde in \cite[Section II]{Wilde-sc}, where
several its important properties have been established.

If $\Phi$ is a quantum operation from $A$ to $B$ then the map $\Phi^*:\B(\H_B)\rightarrow\B(\H_A)$ defined by the relation
\begin{equation}\label{dual-r}
\Tr\shs \Phi(\rho)B=\Tr\shs\Phi^*\!(B)\rho\quad\forall B\in\B(\H_B),\; \rho\in\S(\H_A)
\end{equation}
is called \emph{dual} operation to $\Phi$ \cite{B&R,R&S}. If $\Phi$ is a channel acting on quantum states, i.e. a channel in the Schrodinger pucture, then $\Phi^*$ is a channel acting on quantum observables, i.e. a channel in the Heisenberg picture \cite{H-SCI,Wilde}.

The result in \cite{D-A} implies that the trace-norm convergence in (\ref{star+}) is equivalent to the convergence of the sequence $\{\Phi_n(\rho)\}$ to the operator  $\Phi_0(\rho)$ in the weak operator topology provided that $\{\Tr\Phi_n(\rho)\}$ tends to $\Tr\Phi_0(\rho)$. So, by noting that the set $\S(\H_A)$ in (\ref{star+}) can be replaced by its subset consisting of pure states it is easy to show that the strong convergence of a sequence $\{\Phi_n\}$ of quantum operations  to an operation $\Phi_0$  means that
\begin{equation}\label{sc-ed}
w.o.\shs\textup{-}\lim\limits_{n\rightarrow\infty}\Phi_n^*(B)=\Phi_0^*(B)\;\textup{ for all }\;B\in\B(\H_B),
\end{equation}
where $w.o.\shs\textup{-}\lim$ denotes the limit in the weak operator topology in $\B(\H_A)$.

\section{The strong convergence of quantum operations: the generalized Choi-Jamiolkowski isomorphism and the compactness criterion}

In this section we present a revised version of the compactness criterion for families of quantum operations in the strong convergence
topology obtained in \cite{AQC} in which the subtle inaccuracy made in the original formulation is corrected. As mentioned in the Introduction, this inaccuracy did not affect all the applications of this compactness criterion (known to me). We also describe the proof of the compactness criterion more carefully and consider its equivalent form and applications.\smallskip

We begin by formulating the generalized Choi-Jamiolkowski isomorphism (presented in Proposition 3 in \cite{AQC}) and discussing its corollaries.
\smallskip

Let $\H_A$, $\H_B$ and $\H_R$ be infinite-dimensional
separable  Hilbert spaces. For a given faithful (non-degenerate) state
$\sigma$ in
$\S(\H_R)$ with the spectral representation $\,\sigma=\sum_{i=1}^{+\infty}\lambda_{i}|\psi_i\rangle\langle \psi_i|\,$ denote by $\T(\sigma)$ the
subset of the set
$$
\T_{+,1}(\H_R)\doteq\{\rho\in \T_{+}(\H_R)\,|\,\Tr\rho\leq1\}
$$
consisting of all
operators $\rho$ such that $\sum_{i,j}\frac{\langle
\psi_i|\rho|\psi_j\rangle}{\sqrt{\lambda_{i}\lambda_{j}}}|\psi_i\rangle\langle
\psi_j|\leq I_{R}$ (this means that the matrix
$\left\{\frac{\langle
\psi_i|\rho|\psi_j\rangle}{\sqrt{\lambda_{i}\lambda_{j}}}\right\}_{i,j\geq1}$ corresponds to a bounded
positive operator on $\H_R$ in the sense of Theorem 2 in \cite[Section 29]{A&G}
that is majorized by the unit operator $I_R$  w.r.t. the operator order). \smallskip

Let $\F_{\leq
1}(A,B)$ be the set of all quantum operations from $A$ to $B$ \emph{equipped with the strong convergence
topology} (defined in Section 1). Denote the closed subset of  $\F_{\leq1}(A,B)$
consisting of quantum channels by $\F_{=1}(A,B)$.\smallskip

The following proposition  generalizes the Choi-Jamiolkowski isomorphism
(cf.~\cite{Choi,Jam}) to the case of infinite-dimensional quantum channels and operations.

\smallskip

\begin{property}\label{Y-isomorphism} \cite{AQC}
\textit{Let $\tilde{\omega}$ be a pure state
in $\S(\H_A\otimes\H_R)$ such that
$\,\tilde{\omega}_A\doteq\Tr_{R}\tilde{\omega}$ and $\,\tilde{\omega}_R\doteq\Tr_{A}\tilde{\omega}$ are
faithful states in $\S(\H_A)$ and $\S(\H_R)$, respectively.\footnote{In  Proposition 3 in \cite{AQC}
the faithfulness condition on the state $\tilde{\omega}_A$ is not formulated but used in its proof.} Then the map
\begin{equation}\label{Y-map}
\mathfrak{Y}:\Phi\mapsto \Phi\otimes
\id_R(\tilde{\omega})
\end{equation}
is a topological isomorphism from $\,\F_{\leq
1}(A,B)$ onto the subset
$$
\T_{\leq
1}(\tilde{\omega})\doteq
\left\{\omega\in\T_{+}(\H_B\otimes\H_R)\,|\,\omega_R\in
\T(\tilde{\omega}_R)\right\},
$$
where $\T(\tilde{\omega}_R)$ is the subset of $\,\T_{+}(\H_R)$ defined before.}\smallskip

\textit{The restriction of the map $\mathfrak{Y}$ to the set
$\,\F_{=1}(A,B)$  is a
topological isomorphism from
$\F_{=1}(A,B)$ onto the subset}
$$
\T_{=1}(\tilde{\omega})\doteq
\{\omega\in\S(\H_B\otimes\H_R)\,|\,\omega_R=\tilde{\omega}_R\}.
$$

\emph{The rank of the operator $\Phi\otimes
\id_R(\tilde{\omega})$ is equal to the Choi rank of the operation $\Phi$.}
\end{property}
\smallskip

The last claim of Proposition \ref{Y-isomorphism} follows from its proof presented in \cite{AQC}, where it is shown that any decomposition
of $\Phi\otimes
\id_R(\tilde{\omega})$ into a convex mixture of pure states corresponds to some Kraus representation of $\Phi$. \smallskip

\textbf{Note A:} The closedness of the subset $\T_{\leq
1}(\tilde{\omega})$ of $\T_{+}(\H_B\otimes\H_R)$ in the trace norm follows from
the closedness of the subset $\T(\tilde{\omega}_R)$ of $\T_{+}(\H_R)$ which is shown within the proof of Corollary \ref{aqc-cr}
below (in \cite{AQC} it is stated without proof).\smallskip

\textbf{Note B:} The continuity of the map in $\mathfrak{Y}$ in (\ref{Y-map}) is referred in \cite{AQC} as "an obvious fact", although its proof
requires some efforts.  This claim follows from Proposition 1
in \cite{Wilde-sc}, where the preservation of the strong convergence under the tensor products is established.\smallskip

\begin{remark}\label{note} It is essential that the map $\mathfrak{Y}$ in (\ref{Y-map}) is affine. So, Proposition \ref{Y-isomorphism} shows
that the sets $\,\F_{\leq
1}(A,B)$ and $\T_{\leq
1}(\tilde{\omega})$ (resp. $\,\F_{=1}(A,B)$ and $\T_{=1}(\tilde{\omega})$) are isomorphic as "convex topological spaces".
Among others, this allows us to prove that the sets $\,\F_{\leq 1}(A,B)$, $\,\F_{=1}(A,B)$ and their closed subsets are $\mu$-compact, since any closed subset of $\T_{+,1}(\H_B\otimes\H_R)$ is $\mu$-compact \cite{P&Sh}.\footnote{The $\mu$-compactness is a property of a subset
of a topological linear space reflecting a special relation between the topology and the structure of linear operations. It can be treated as a weakened form of compactness, since
\begin{itemize}
  \item any compact subset of a topological linear space is $\mu$-compact;
  \item many well known results valid for compact convex sets are generalized to convex $\mu$-compact sets (in particular, the Krein-Milman theorem and some results of the Choque theory) \cite{P&Sh}.
\end{itemize}}
It follows, in particular, that the Krein-Milman theorem is valid for the non-compact sets $\,\F_{\leq 1}(A,B)$, $\,\F_{=1}(A,B)$ and their closed convex subsets (due to Proposition 5 in \cite{P&Sh}).
\end{remark}

\begin{remark}\label{note+}
Another benefit of the isomorphism  $\mathfrak{Y}$ is a simple proof of the closedness of the set of quantum operations (channels)
with the Choi rank not exceeding a given $n\in\N$  w.r.t. the strong convergence (due to the last claim of Proposition \ref{Y-isomorphism}).
\end{remark}
\smallskip

The following corollary contains a revised version of the
compactness criterion for families  of quantum operations
in the strong convergence topology.\smallskip

\begin{corollary}\label{aqc-cr}
A) \emph{A \textbf{closed} subset
$\F_{0}\subseteq\F_{\leq
1}(A,B)$ is compact if and only if there exists a faithful
state $\sigma$  in $\S(\H_A)$ such that
$\{\Phi(\sigma)\}_{\Phi\in\F_{0}}$ is a compact subset
of $\,\T_{+}(\H_B)$.}\smallskip

B) \emph{If $\,\F_{0}$ is a compact subset of $\,\F_{\leq
1}(A,B)$ then
$\{\Phi(\sigma)\}_{\Phi\in\F_{0}}$ is a compact subset
of $\T_{+}(\H_B)$ for arbitrary state $\sigma$
in $\S(\H_A)$.}\smallskip
\end{corollary}

Corollary \ref{aqc-cr} is valid with $\,\F_{\leq
1}(A,B)$ and $\T_{+}(\H_B)$ replaced by $\,\F_{=1}(A,B)$ and $\S(\H_B)$.

 \smallskip

\textit{Proof.} A) Assume that $\,\sigma=\sum_{i}\lambda_{i}|\varphi_i\rangle\langle \varphi_i|\,$ is a faithful state in $\S(\H_A)$ such that
$\{\Phi(\sigma)\}_{\Phi\in\F_{0}}$ is a compact subset
of $\T_{+}(\H_B)$ (here $\{\varphi_i\}$ is an orthonormal basis in $\H_A$). Then there is a pure state $\tilde{\omega}$
in $\S(\H_A\otimes\H_R)$ such that
$\,\tilde{\omega}_A=\sigma$ and $\,\tilde{\omega}_R$ is a
faithful state in $\S(\H_R)$ with the spectral representation $\tilde{\omega}_R=\sum_{i}\lambda_{i}|\psi_i\rangle\langle \psi_i|$, where $\{\psi_i\}$ is an orthonormal basis in $\H_R$ \cite{H-SCI,Wilde}.

Show first that the set $\mathfrak{T}(\tilde{\omega}_R)$ (defined before Proposition \ref{Y-isomorphism}) is a
compact subset of $\mathfrak{T}_{+}(\H_R)$. The relative compactness
of $\mathfrak{T}(\tilde{\omega}_R)$ follows from
the compactness criterion for subsets of
$\mathfrak{T}_{+}(\H_R)$ \cite[Proposition 11]{AQC}.
Indeed, if $P_{n}=\sum_{i=1}^{n}|\psi_i\rangle\langle \psi_i|$ then the definition of $\mathfrak{T}(\tilde{\omega}_R)$  implies
$$
\Tr\rho(I_R-P_{n})=\sum_{i>n}\langle
\psi_i|\rho|\psi_i\rangle\leq \sum_{i>n}\lambda_{i},\quad\forall
\rho\in\T(\tilde{\omega}_R).
$$
The closedness of $\mathfrak{T}(\tilde{\omega}_R)$ can be derived from Theorem 2 in \cite{A&G}. To give an explicit proof assume that
$\{\rho_n\}$ is a sequence of operators in $\mathfrak{T}(\tilde{\omega}_R)$ converging
to an operator $\rho_0$. Let $A_n$ be the bounded positive operator on $\H_R$ determined by the matrix $\left\{\frac{\langle
\psi_i|\rho_n|\psi_j\rangle}{\sqrt{\lambda_{i}\lambda_{j}}}\right\}_{i,j}$ in the basic $\{\psi_i\}$. Since $A_n\leq I_R$ for all $n$,
the compactness of the unit ball of $\B(\H_R)$ in the weak operator topology (cf.\cite{B&R}) implies the existence of a subsequence  $\{A_{n_k}\}$
weakly converging to a positive operator $A_0\leq I_R$.\footnote{We use the fact that the  weak operator topology on the unit ball of $\B(\H_R)$ is metrizable \cite{B&R}.} It is easy to see that $\left\{\frac{\langle
\psi_i|\rho_0|\psi_j\rangle}{\sqrt{\lambda_{i}\lambda_{j}}}\right\}_{i,j}$ is the matrix of $A_0$ in the basic $\{\psi_i\}$. Hence, $\rho_0$
belongs to the set $\mathfrak{T}(\tilde{\omega}_R)$.\smallskip

The compactness of the sets $\{\Phi(\sigma)\}_{\Phi\in\F_{0}}$ and $\mathfrak{T}(\tilde{\omega}_R)$ implies, by Corollary 6 in \cite[the Appendix]{AQC}, that
the set $\{\Phi\otimes
\id_R(\tilde{\omega})\}_{\Phi\in\mathfrak{F}_{0}}$ is relatively compact. So, the compactness of the \emph{closed} set $\mathfrak{F}_{0}$ in the
strong convergence topology  follows from  Proposition \ref{Y-isomorphism}.\smallskip

B) Since the compactness is preserved under the action of continuous maps, this assertion obviously follows from the definition of the
strong convergence topology. $\square$\smallskip

\begin{remark} The assumption of closedness of the set $\F_0$ in Corollary \ref{aqc-cr}A is essential.
To show this take any sequence $\{\sigma_n\}$ in $\S(\H_A)$ converging to a faithful state $\sigma_0$  and consider the countable set $\F_*=\{\Phi_n\}_{n\geq0}$ of quantum channels from $A$ to $B=A$,
where $\Phi_0=\id_A$ is the identity channel and $\Phi_n(\rho)=[\Tr\rho]\sigma_n$ for $n>0$. Then $\{\Phi(\sigma_0)\}_{\Phi\in\F_*}$ is the compact set $\{\sigma_n\}_{n\geq0}$. Nevertheless,
the set $\F_*$ is not compact in the strong convergence topology, since it is not closed: the sequence $\{\Phi_n\}_{n>0}$ strongly converges
to the channel $\Phi_*(\rho)=[\Tr\rho]\sigma_0$ not belonging to the set $\F_*$. This shows the necessity to correct
the statement of Corollary 2A in \cite{AQC}.
\end{remark}\smallskip

\begin{remark} The proof of part A of  Corollary \ref{aqc-cr}  based on the generalized Choi-Jamiolkowski isomorphism
(presented in Proposition \ref{Y-isomorphism}) is simple but it does not explain how the compactness of the set $\,\{\Phi(\sigma)\}_{\Phi\in\F_{0}}$
for only one faithful state $\sigma$ implies the compactness of $\F_{0}$ (which, in turn, implies the compactness of the set $\,\{\Phi(\sigma)\}_{\Phi\in\F_{0}}$ for all states $\sigma$ by part B of  Corollary \ref{aqc-cr}).

To clarify this point one can give a direct proof of Corollary \ref{aqc-cr} based on the compactness criterion for
bounded sets of positive trace class operators \emph{which does not use the complete positivity} of the maps in $\F_{0}$.
Proposition \ref{comp-c} in the Appendix contains a
compactness criterion for norm bounded families of positive linear maps between the Banach spaces $\T(\H_A)$ and $\T(\H_B)$
in the strong convergence topology  which looks very similar to the compactness criterion in Corollary \ref{aqc-cr}. The proof of this proposition can be treated as a direct proof of Corollary \ref{aqc-cr}, since the set of all quantum operations (resp. channels) from $A$ to $B$ is a closed subset of the set of all
trace-non-increasing (resp. trace preserving) positive  linear maps between $\T(\H_A)$ and $\T(\H_B)$ w.r.t. the strong convergence.
\end{remark}\smallskip

The arguments used in the proof of Corollary \ref{aqc-cr} allow us to obtain
its following modification.\smallskip

\begin{corollary}\label{aqc-cr-c}
A) \emph{A subset
$\,\F_{0}\subseteq\F_{\leq
1}(A,B)$ is relatively compact if and only if there exists a faithful
state $\sigma$  in $\S(\H_A)$ such that
$\,\{\Phi(\sigma)\}_{\Phi\in\F_{0}}$ is a relatively compact subset
of $\,\T_{+}(\H_B)$.}\smallskip

B) \emph{If $\,\F_{0}$ is a relatively compact subset of $\,\F_{\leq
1}(A,B)$ then
$\{\Phi(\sigma)\}_{\Phi\in\F_{0}}$ is a relatively compact subset
of $\,\T_{+}(\H_B)$ for arbitrary state $\sigma$
in $\S(\H_A)$.}
\end{corollary}
\medskip
Corollary \ref{aqc-cr-c} is valid with $\,\F_{\leq
1}(A,B)$ and $\T_{+}(\H_B)$ replaced, respectively,  by $\,\F_{=1}(A,B)$ and $\S(\H_B)$.

\smallskip

In a sense, the compactness criterion in the form of Corollary \ref{aqc-cr-c} looks more
natural, since its part A does not contain additional assumptions about the set $\F_{0}$.\smallskip

\section{Applications}

\subsection{Basic lemmas with illustrating examples}

Concrete applications of the results of Section 2 are often based on the following two lemmas
proved by using Corollary \ref{aqc-cr-c}.\smallskip

\begin{lemma}\label{l-1}  \emph{Let $\{\rho_n\}\subset\T_+(\H_A)$ be a sequence converging to a faithful state
$\rho_0$ in $\S(\H_A)$ and $\{\Phi_n\}$ be a sequence of quantum operations from $A$ to $B$ such that
\begin{equation}\label{l-1-1}
\lim_{n\to+\infty}\Phi_n(\rho_n)=\sigma_0\in\T_+(\H_B).
\end{equation}
Then the sequence $\{\Phi_n\}$ is relatively compact in the strong convergence topology and any its partial
limit $\Phi_0$ has the properties
\begin{equation}\label{l-1-2}
\Phi_0(\rho_0)=\sigma_0,\qquad  \Ch(\Phi_0)\leq \limsup_{n\to+\infty}\Ch(\Phi_n)
\end{equation}
where $\Ch(\Psi)$ denotes the Choi rank of an operation $\Psi$.}
\end{lemma}\smallskip

\emph{Proof.} To derive the main claim of Lemma \ref{l-1} from Corollary \ref{aqc-cr-c} it suffices to note that its
conditions imply that (\ref{l-1-1}) holds with $\rho_n$ replaced by $\rho_0$ (due to the uniform boundedness of the operator norms
of all the maps $\Phi_n$).

The first relation in (\ref{l-1-2}) is obvious, the second one follows from the closedness of the set of quantum operations
with the Choi rank not exceeding a given bound (see  Remark \ref{note+} in Section 2).  $\Box$\smallskip

\begin{example}\label{l-1-e} Let $\{\rho_n\}\subset\T_+(\H)$ be a sequence converging to a faithful state
$\rho_0$ in $\S(\H)$ and $\{A_n\}$ be a sequence of operators from the unit ball of $\B(\H)$ such that
\begin{equation*}
\lim_{n\to+\infty}A_n\rho_n A^*_n=\sigma_0\in\T_+(\H).
\end{equation*}
By Lemma \ref{l-1} (along with  Lemma \ref{1-rank} in Section 3.2.5 below) the sequence $\{A_n\}$ is relatively compact in the strong operator topology and any its partial
limit $A_0$ has the property $A_0\rho_0 A^*_0=\sigma_0$.
\end{example}\smallskip

\begin{lemma}\label{l-2}  \emph{Let $\{\Phi_n\}$ and $\{\Psi_n\}$ be sequences of quantum operations from $A$ to $B$
such that
\begin{itemize}
  \item $\{\Phi_n\}$ strongly converges to an operation $\Phi_0$;
  \item there is a faithful state $\sigma$ in $\S(\H_A)$ and $c>0$ s.t. $\,c\Psi_n(\sigma)\leq\Phi_n(\sigma)$ for all $n\neq0$.
\end{itemize}
Then  the sequence $\{\Psi_n\}$ is relatively compact in the strong convergence topology and any its partial
limit $\Psi_0$ has the properties
\begin{equation}\label{l-2-1}
c\Psi_0(\sigma)\leq\Phi_0(\sigma),\qquad  \Ch(\Psi_0)\leq \limsup_{n\to+\infty}\Ch(\Psi_n),
\end{equation}
where $\Ch(\Psi)$ denotes the Choi rank of an operation $\Psi$.}
\end{lemma}\smallskip

\emph{Proof.} The strong convergence of $\Phi_n$ to $\Phi_0$ implies that the set $\{\Phi_n(\sigma)\}_{n\geq0}$ is compact.
Thus, the relation $c\Psi_n(\sigma)\leq\Phi_n(\sigma)$ allows us to show (by using the compactness criterion from Proposition 11 in \cite[the Appendix]{AQC})
that the set $\{\Psi_n(\sigma)\}_{n>0}$ is relatively compact. So, the main claim of Lemma \ref{l-2} follows from Corollary \ref{aqc-cr-c} in Section 2.

The first relation in (\ref{l-2-1}) is obvious, the second one follows from the closedness of the set of quantum operations
with the Choi rank not exceeding a given bound (Remark \ref{note+} in Section 2).  $\Box$\smallskip

\begin{example}\label{l-2-e} Let $\{\Phi_n\}$ be a sequence of quantum channels from $A$ to $B$
strongly converging to a channel $\Phi_0$. Assume that $\Ch(\Phi_n)\leq m$ and
$\Phi_n(\rho)=\sum_{i=1}^m A_i^n\rho\shs[A_i^n]^*$ is the Kraus representation of $\Phi_n$ for any $n$.

Lemma \ref{l-2} implies that
the sequence of quantum operations $\Psi^i_n(\rho)=A_i^n\rho\shs[A_i^n]^*$ is relatively compact in the strong convergence topology for each $i$
and that all its partial limits are operations with Choi rank $\leq 1$.  By using this and Lemma \ref{1-rank} in Section 3.2.5 below it is easy to show the existence
of operators $A^0_1$,...,$A^0_m$ and an increasing sequence $\{n_k\}$ of natural numbers such that
$$
s.o.\textup{-}\lim_{k\to+\infty}A_i^{n_k}=A_i^{0}\quad\forall i\quad \textrm{and} \quad\Phi_0(\rho)=\sum_{i=1}^m A_i^0\rho\shs[A_i^0]^*,\quad \rho\in\S(\H_A),
$$
where $s.o.\textup{-}\lim$ denotes the limit in the strong operator topology. This means, roughly speaking, that \emph{from any sequence of Kraus representations of a strongly converging sequence
of quantum channels with bounded Choi rank it is possible to extract a subsequence converging to the Kraus representation of a limit channel}.

It is not hard to construct an example showing that the above claim is not valid without the condition $\sup_n\Ch(\Phi_n)<+\infty$.
\end{example}

\subsection{Simple applications}

\subsubsection{The set of quantum operations (resp. channels) mapping a given input state into a  given output operator (resp. state)}

Let $\sigma$ be a faithful state in $\mathfrak{S}(\mathcal{H}_A)$
and $\rho$ be an arbitrary positive operator in the unit ball of
$\mathfrak{T}(\mathcal{H}_B)$. By Corollary \ref{aqc-cr}
the set
$$
\mathfrak{F}_{\leq
1}^{\,\sigma\mapsto \rho}=\{\Phi\in\mathfrak{F}_{\leq
1}(A,B)\,|\,\Phi(\sigma)=\rho\}
$$
of all quantum operations mapping the state $\sigma$ into the operator $\rho$
is compact in the strong convergence topology.  Note that this
set is not compact in the topology of uniform convergence. Note
also that the set of \textit{all} CP linear maps transforming the state
$\sigma$ into a given operator $\rho$ is not compact in the
strong convergence topology.

By Proposition \ref{Y-isomorphism} the set $\mathfrak{F}_{\leq
1}^{\,\sigma\mapsto \rho}$ is topologically and affinely isomorphic to the closed convex subset
$\left\{\omega\in\T_{+}(\H_B\otimes\H_R)\,|\,\omega_B=\rho,\omega_R\in
\T(\tilde{\sigma})\right\}$
of $\T_{+}(\H_B\otimes\H_R)$, where $\tilde{\sigma}$ is a state of a reference system $R\cong A$ unitary equivalent to $\sigma$
and $\T(\tilde{\sigma})$ is the closed convex subset of $\T_+(\H_R)$ defined via $\tilde{\sigma}$ by the rule described before Proposition \ref{Y-isomorphism}.
\smallskip

If $\rho$ is a state in $\mathfrak{S}(\mathcal{H}_B)$
then the closedness of $\mathfrak{F}_{=1}(A,B)$ in $\mathfrak{F}_{\leq1}(A,B)$ shows that
the set
$$
\mathfrak{F}_{=1}^{\,\sigma\mapsto \rho}=\{\Phi\in\mathfrak{F}_{=1}(A,B)\,|\,\Phi(\sigma)=\rho\}
$$
of all quantum channels mapping the state $\sigma$ into the state $\rho$
is compact in the strong convergence topology. Moreover, by Proposition \ref{Y-isomorphism} the set $\mathfrak{F}_{=1}^{\,\sigma\mapsto \rho}$ is topologically and affinely isomorphic to the closed convex subset $\left\{\omega\in\T_{+}(\H_B\otimes\H_R)\,|\,\omega_B=\rho,\omega_R=\tilde{\sigma}\right\}$
of $\T_{+}(\H_B\otimes\H_R)$, where $\tilde{\sigma}$ is a given state of a reference system $R\cong A$ unitary equivalent to $\sigma$.
\smallskip

The above claim implies, in particular, that an arbitrary family $\F_0$ of quantum channels
having a faithful invariant state $\sigma$ (i.e. such that $\Phi(\sigma)=\sigma$ for all $\Phi\in\F_0$)
is relatively compact in the strong convergence topology.  \smallskip

\subsubsection{The set of channels with bounded energy amplification factor}

Let $\sigma$ be a faithful state in $\mathfrak{S}(\mathcal{H}_A)$ and
$H_B$ be a positive unbounded operator on $\H_B$ with a discrete spectrum of finite multiplicity, which
can be interpreted as a Hamiltonian of a quantum system
described by the space $\mathcal{H}_B$. Corollary
\ref{aqc-cr} implies that the set
$$
\left\{\Phi\in\mathfrak{F}_{=1}(A,B)\,|\,\mathrm{Tr}H_B\Phi(\sigma)\leq
E\shs\right\}
$$
is compact in the strong convergence topology for each $E>0$, since this set is closed w.r.t. the strong convergence due to the lower semicontinuity
of the function $\rho\mapsto \Tr H_B\rho\,$ and
the subset $\{\rho\in\S(\H_B)\,|\,\Tr H_B\rho\leq E\}$ is compact by the Lemma in \cite{H-c-w-c}.\smallskip

Let $H_A$ be a densely defined positive operator on
$\mathcal{H}_A$. For given $K>0$ consider the set
\begin{equation*}
\mathfrak{F}_{H_A,H_B,K}=\left\{\Phi\in\mathfrak{F}_{=1}(A,B)\,\left|\,
\sup_{\rho\in\mathfrak{S}(\mathcal{H}_A),\mathrm{Tr}H_A\rho<+\infty}
\frac{\mathrm{Tr}H_B\Phi(\rho)}{\mathrm{Tr}H_A\rho}\leq
K\right.\right\}
\end{equation*}

If  $H_A$ and $H_B$ are
Hamiltonians of the systems $A$ and $B$, respectively, then
$\mathfrak{F}_{H_A,H_B,K}$ is the set of channels from $A$ to $B$
with the energy amplification factor not exceeding  $K$. By the
above observation the set  $\mathfrak{F}_{H_A,H_B,K}$ is compact in
the strong convergence topology for each $K$.

\subsubsection{Criteria for the existence of a limit point of a sequence of quantum operations w.r.t. the strong convergence}

Practical applications of the compactness criterion presented in Section 2  often consist in proving the
existence of a limit point (partial limit) for a given sequence of quantum channels or operations (it is this trick that is used in almost all applications considered in Section 3 below).
In the following proposition we collect several criteria for the existence of a limit point of a sequence of quantum operations (or channels) obtained by using Corollary \ref{aqc-cr-c} in Section 2.\smallskip

\begin{property}\label{new-cor}
\emph{Let $\{\Phi_n\}_{n\in\N}$ be a sequence of quantum operations from $A$ to $B$ and $\{\Phi^*_n\}_{n\in\N}$ the corresponding sequence of dual operations (defined in (\ref{dual-r})). The following properties are equivalent:}
\begin{enumerate}[(i)]
  \item \emph{the sequence $\{\Phi_n\}_{n\in\N}$ has a limit point in $\F_{\leq1}(A,B)$;}
  \item \emph{there is a faithful state $\sigma$ in $\S(\H_A)$ such that the sequence $\{\Phi_n(\sigma)\}_{n\in\N}$ has a limit point in $\T_+(\H_B)$;}
  \item \emph{there exist a faithful state $\sigma$ in $\S(\H_A)$, an increasing sequence $\{n_k\}_{k\in\N}$  of natural numbers and an increasing sequence $\{P_m\}_{m\in\N}$ of finite-rank projectors in $\B(\H_B)$ converging to the unit operator $I_B$ in the strong operator topology  such that}
      $$
      \lim_{m\to+\infty}\sup_{k\in\N}\Tr (I_B-P_m)\Phi_{n_k}(\sigma)=0;
      $$
  \item \emph{there exist  an increasing sequence $\{P_m\}_{m\in\N}$ of finite-rank projectors in $\B(\H_B)$ converging to the unit operator $I_B$ in the strong operator topology,  an increasing sequence $\{n_k\}_{k\in\N}$  of natural numbers and a  sequence $\{A_m\}\subset\B_+(\H_A)$ such that
\begin{equation}\label{s-c-1}
   w.o.\shs\textup{-}\lim_{k\to+\infty} \Phi_{n_k}^*(P_m)=A_m\quad\forall m\quad\;\; \textit{and}\quad\;\;   w.o.\shs\textup{-}\lim_{k\to+\infty} \Phi_{n_k}^*(I_B)=\sup_{m\in\N}A_m,
\end{equation}
where $w.o.\shs\textup{-}\lim$ denotes the limit in the weak operator topology in $\B(\H_A)$ and $\sup_{m\in\N}A_m$ is the least upper bound of the nondecreasing sequence $\{A_m\}$ \cite{B&R,H-SSQT};}
     \item \emph{there exist a faithful state $\sigma$ in $\S(\H_A)$, an increasing sequence $\{P_m\}_{m\in\N}$ of finite-rank projectors in $\B(\H_B)$ converging to the unit operator $I_B$ in the strong operator topology
   and  an increasing sequence $\{n_k\}_{k\in\N}$  of natural numbers such that}
\begin{equation}\label{s-c-2}
      \lim_{k\to+\infty} \Tr\Phi_{n_k}^*(P_m)\sigma=a_m\in\mathbb{R}_+\quad\forall m\quad\;\; \textit{and}\quad\;\;   \lim_{k\to+\infty}\Tr \Phi_{n_k}^*(I_B)\sigma=\sup_{m\in\N}a_m.
\end{equation}
\end{enumerate}

\emph{If $\{\Phi_n\}_{n\in\N}$ is a sequence of quantum channels then
\begin{itemize}
  \item $\F_{\leq1}(A,B)$ in $(i)$ and $\T_+(\H_B)$ in $(ii)$ are replaced by $\F_{=1}(A,B)$ and $\S(\H_B)$;
  \item the second conditions in (\ref{s-c-1}) and (\ref{s-c-2})
means, respectively,  that $\,\sup_{m\in\N}A_m=I_A$ and $\;\sup_{m\in\N}a_m=1$.
\end{itemize}}
\end{property}\medskip

\emph{Proof.} The  implication $\rm (i)\Rightarrow(ii)$ is obvious. Corollary \ref{aqc-cr-c} proves the implication $\rm (ii)\Rightarrow(i)$,
since the existence of a limit point of the sequence $\{\Phi_n(\sigma)\}_{n\in\N}$ is equivalent to the existence of a relatively compact subsequence of this sequence.
The  implication $\rm (iii)\Rightarrow(ii)$ follows  from the compactness criterion for bounded sets of positive
trace class operators \cite[Proposition 11]{AQC}. The  implication $\rm (iv)\Rightarrow(v)$ follows from the coincidence
of the weak operator topology and the $\sigma$-weak (ultra-weak) operator topology on the unit ball of $\B(\H_A)$ \cite{B&R}.

Thus, we have to prove the  implications $\rm (i)\Rightarrow(iv)$ and  $\rm (v)\Rightarrow(iii)$.

If $\rm (i)$ holds  then there is an increasing sequence $\{n_k\}_{k\in\N}$  of natural numbers such that the sequence $\{\Phi_{n_k}\}_k$
strongly converges to a quantum operation $\Phi_0$. This implies, due to characterization (\ref{sc-ed}) of the strong convergence, that
$$
w.o.\shs\textup{-}\lim_{k\to+\infty} \Phi_{n_k}^*(P_m)=\Phi_{0}^*(P_m)\quad\forall m\quad\;
\; \textup{and}\quad\;\;   w.o.\shs\textup{-}\lim_{k\to+\infty} \Phi_{n_k}^*(I_B)=\Phi_{0}^*(I_B)
$$
for any increasing sequence $\{P_m\}_{m\in\N}$ of finite-rank projectors in $\B(\H_B)$ strongly converging to the unit operator $I_B$.
Since the dual operation $\Phi_{0}^*$ to the operation $\Phi_0$ is a normal\footnote{A map $\Psi:\B(\H)\to\B(\H')$ is called normal if $\Psi(\sup_{\lambda}A_\lambda)=\sup_{\lambda}\Psi(A_{\lambda})$ for any increasing net $A_\lambda\subset\B(\H)$ \cite{B&R,H-SSQT}.}  map \cite{B&R,H-SSQT}, we have $\Phi_{0}^*(I_B)=\sup_{m\in\N}\Phi_{0}^*(P_m)$. Thus, $\rm (iv)$
is valid with  $A_m=\Phi_{0}^*(P_m)$.

Assume that $\rm (v)$ holds. Since $\Tr\Phi_{n_k}^*(P_m)\sigma\leq \Tr\Phi_{n_k}^*(P_{m+1})\sigma$  and $a_m\leq a_{m+1}$ for all $m$ and $k$, by using Dini's lemma  it is easy to show that $\Tr\Phi_{n_k}^*(P_m)\sigma$ tends to $\Tr\Phi_{n_k}^*(I_B)\sigma$ as $m\to+\infty$ uniformly on $k$. This implies $\rm (iii)$. \smallskip

The last claim of the proposition is obvious. $\Box$
\medskip

\begin{remark}\label{int} If $\{\Phi_n\}_{n\in\N}$ is an arbitrary sequence of quantum operations from $A$ to $B$ then by using the compactness of the unit ball of $\B(\H_A)$
in the weak operator topology (cf.\cite{B&R}) and the "diagonal" method one can show that
for any given increasing sequence $\{P_m\}_{m\in\N}$ of finite-rank projectors in $\B(\H_B)$ strongly converging to the unit operator $I_B$ there is an increasing sequence $\{n_k\}_{k\in\N}$  of natural numbers such that
\begin{equation}\label{int-rel}
 w.o.\shs\textup{-}\lim_{k\to+\infty} \Phi_{n_k}^*(P_m)=A_m\quad\forall m\quad\;\; \textup{and}\quad\;\;   w.o.\shs\textup{-}\lim_{k\to+\infty} \Phi_{n_k}^*(I_B)=A_*,
\end{equation}
where $A_m$ and $A_*$ are positive operators in $\B(\H_A)$. So, a critical point of property $\rm (iv)$ in Proposition \ref{new-cor} is the coincidence of
$\sup_m A_m$ and $A_*$. If $\sup_m A_m=A_*$ for at least one sequence  $\{P_m\}_{m\in\N}$
then the sequence  $\{\Phi_n\}_{n\in\N}$  contains a strongly converging subsequence by Proposition \ref{new-cor} and hence property $\rm (iv)$ holds for any sequence  $\{P_m\}_{m\in\N}$ by the
proof the  implication $\rm (i)\Rightarrow(iv)$.
In the general case, we have $\sup_m A_m\leq A_*$.

To illustrate the above observations consider the sequence of channels
$$
\Phi_n(\rho)=V_n\rho V_n^*,\quad \rho\in\S(\H_A),
$$
determined by a sequence $\{V_n\}_{n\in\N}$ of  isometries from $\H_A$ to $\H_B$ with mutually orthogonal ranges (i.e. such that
$V^*_iV_j=0$ for all $i\neq j$). It is clear that the  sequence $\{\Phi_n\}_{n\in\N}$ has no limit points w.r.t. the strong convergence. Since
$\Phi^*_n(B)=V^*_nBV_n$, it is easy to see that for any given increasing sequence $\{P_m\}_{m\in\N}$ of
finite-rank projectors in $\B(\H_B)$ strongly converging to the unit operator $I_B$ the limit relations in
(\ref{int-rel}) hold with $n_k=k$, $A_m=0$ and $A_*=I_A$.
\end{remark}\smallskip

\begin{remark}\label{two} It is mentioned at the end of Section 1 that the strong convergence
of a sequence $\{\Phi_n\}_{n\in\N}$ of quantum channels from $A$ to $B$ to a quantum channel $\Phi_0$ means that
\begin{equation}\label{sc-ed+}
w.o.\shs\textup{-}\lim\limits_{n\rightarrow\infty}\Phi_n^*(B)=\Phi_0^*(B)
\end{equation}
for any $B\in\B(\H_B)$. At the same time, to ensure the existence of limit points of
the sequence $\{\Phi_n\}_{n\in\N}$ it suffices, by Proposition \ref{new-cor}, to check the validity of (\ref{sc-ed+})
when $B$ runs over a particular increasing sequence $\{P_m\}_{m\in\N}$
of finite-rank projectors in $\B(\H_B)$ converging to $I_B$ in the strong operator topology. Indeed, in this case $\rm (iv)$
holds with $A_m=\Phi_0^*(P_m)$ because $\,I_A=\Phi_{n}^*(I_B)=\Phi_{0}^*(I_B)=\sup_{m\in\N}\Phi_{0}^*(P_m)\,$ due to the normality of the map $\Phi_{0}^*$ \cite{B&R,H-SSQT}.

Although the validity of (\ref{sc-ed+}) with $B=P_m$ for all $m$ implies the existence of limit points of $\{\Phi_n\}_{n\in\N}$ w.r.t  the strong convergence, it does not
imply that $\Phi_0$ is a limit point of this sequence. This can be
illustrated by the following example.

Let $\{\{\varphi_k^n\}_{k\in\N}\}_n$ be a sequence of
orthonormal base in a separable Hilbert space $\H_A$ converging to an orthonormal basis $\{\varphi_k^0\}_{k\in\N}$ in $\H_A$ in the sense that
$\varphi_k^n$ tends to $\varphi_k^0$ as $n\to+\infty$ for all $k$. Consider the sequence of channels
$$
\Phi_n(\rho)=\sum_{k=1}^{+\infty} \langle\varphi^n_k|\rho|\varphi^n_k\rangle|\varphi^0_k\rangle\langle\varphi^0_k|, \quad \rho\in\S(\H_A)
$$
from $A$ to $B=A$ and the channel $\Phi_0=\id_A$. Let $P_m=\sum_{k=1}^{m}|\varphi^0_k\rangle\langle\varphi^0_k|$ for all $m\in\N$. Then
the sequence $\{P_m\}_{m\in\N}$ consists of finite-rank projectors and strongly converges to $I_B$. We  have
$$
\Phi^*_n(P_m)=\sum_{k=1}^{+\infty} \langle\varphi^0_k|P_m|\varphi^0_k\rangle|\varphi^n_k\rangle\langle\varphi^n_k|=\sum_{k=1}^{m}|\varphi^n_k\rangle\langle\varphi^n_k|.
$$
Thus, $\Phi^*_n(P_m)$ tends to $\Phi^*_0(P_m)=P_m$ in the operator norm for each $m$, which shows that property $\rm (iv)$ in Proposition \ref{new-cor} holds. But, the channel $\Phi_0$
is not a limit point of the sequence $\{\Phi_n\}_{n\in\N}$, since this sequence strongly converges to the
channel
$$
\Psi(\rho)=\sum_{k=1}^{+\infty} \langle\varphi^0_k|\rho|\varphi^0_k\rangle|\varphi^0_k\rangle\langle\varphi^0_k|, \quad \rho\in\S(\H_A).
$$
So, the sequence $\{\Phi_n\}_{n\in\N}$ has a unique limit point, which does not coincide with $\Phi_0$.\smallskip

The family $\{\Phi_n\}_{n\geq0}$ is another example showing that the condition of closedness of the family $\F_0$
in part A of Corollary \ref{aqc-cr} in Section 2 is  necessary. Indeed, it is easy to see that $\{\Phi_n(\sigma)\}_{n\geq0}$
is a compact subset of $\T(\H_B)$ for any faithful state $\sigma$ in $\H_A$ diagonizable in the basis $\{\varphi_k^0\}_{k\in\N}$.
\end{remark}

\subsubsection{Proof of the strong convergence for specific sequences of  quantum channels and operations}

A criterion for the relative compactness of families of quantum channels and operations considered in Section 2 gives an additional way
to prove the strong convergence of  sequences of quantum channels and operations. Indeed, to prove that a sequence $\{\Phi_n\}$ of quantum operations
from $A$ to $B$ strongly converges to a quantum operation $\Phi_0$ it suffices to prove its relative compactness and to show that all the limit
points of this sequence coincide with $\Phi_0$. By Corollary \ref{aqc-cr-c} this can be done in two steps:
  \begin{enumerate}[1)]
    \item  find a faithful state $\sigma$ in $\S(\H_A)$ such that the sequence $\{\Phi_n(\sigma)\}$ is relatively compact (in particular,
    is converging to the operator $\Phi_0(\sigma)$);
    \item assuming that a subsequence $\{\Phi_{n_k}\}$ strongly converges to an operation $\Theta$  show that $\Theta=\Phi_0$.
  \end{enumerate}

\begin{example}\label{Wilde-exam}
Proposition 1 in \cite{Wilde-sc} states that the strong convergence of sequences $\{\Phi_n\}\subset\F_{=1}(A,B)$ and $\{\Psi_n\}\subset\F_{=1}(C,D)$ of quantum channels to
channels $\Phi_0$ and $\Psi_0$ implies the strong convergence of the sequence $\{\Phi_n\otimes\Psi_n\}\subset\F_{=1}(AC,BD)$ to the channel $\Phi_0\otimes\Psi_0$.
The above two-step approach allows us to essentially simplify the proof of this claim. Indeed, for the first step it suffices to take a faithful state $\sigma=\tilde{\alpha}\otimes\tilde{\gamma}$, where
$\tilde{\alpha}$ and $\tilde{\gamma}$ are given faithful states of $A$ and $C$, respectively, since it is clear that $\Phi_n\otimes\Psi_n(\tilde{\alpha}\otimes\tilde{\gamma})$ tends to $\Phi_0\otimes\Psi_0(\tilde{\alpha}\otimes\tilde{\gamma})$. The second step is easily realized: if $\{\Phi_{n_k}\otimes\Psi_{n_k}\}$ is a subsequence strongly converging to a channel $\Theta$ then the relation $$
\Phi_0\otimes\Psi_0(\alpha\otimes\gamma)=\lim_{k\to+\infty}\Phi_{n_k}\otimes\Psi_{n_k}(\alpha\otimes\gamma)=\Theta(\alpha\otimes\gamma)
$$
valid for arbitrary $\alpha\in\T(\H_A)$ and $\gamma\in\T(\H_C)$ implies that $\Theta=\Phi_0\otimes\Psi_0$.
\end{example}

It is essential, that the above arguments remain valid in the case when $\{\Phi_n\}$ and $\{\Psi_n\}$ are
sequences of quantum operations  strongly converging to quantum operations $\Phi_0$ and $\Psi_0$ (this case is not covered by Proposition 1 in \cite{Wilde-sc} and its proof).
The corresponding generalization of Proposition 1 in \cite{Wilde-sc} is used below, so we formulate it as

\begin{lemma}\label{s-c-t-pr} \emph{If $\,\{\Phi_n\}\subset\F_{\leq 1}(A,B)$ and $\,\{\Psi_n\}\subset\F_{\leq 1}(C,D)$ are sequences of quantum operations
strongly converging to quantum operations $\Phi_0$ and $\Psi_0$ then the sequence $\{\Phi_n\otimes\Psi_n\}\subset\F_{\leq 1}(AC,BD)$ strongly converges
to the quantum operation $\Phi_0\otimes\Psi_0$.}
\end{lemma}

\medskip
Another example of using the above two-step approach to prove the strong convergence can be found at the end of Section 3.3.

\subsubsection{Criterion of relative
compactness of bounded subsets of $\B(\H,\K)$ in the strong operator topology}

Let $\B(\H,\K)$ be the space of all bounded linear
operators from a separable Hilbert space $\H$ to a separable Hilbert space $\K$. Corollary
\ref{aqc-cr-c} implies the following criterion of relative
compactness of bounded subsets of  $\B(\H,\K)$ in the strong operator topology.\smallskip\pagebreak

\begin{property}\label{new-com-cr} \emph{Let $\B$ be a bounded subset of $\B(\H,\K)$. Then the following properties
are equivalent:}
\begin{enumerate}[(i)]
  \item \emph{the set $\B$ is relatively compact in the strong operator topology in $\B(\H,\K)$;}
  \item \emph{there is a faithful state $\sigma\in\S(\H)$ such that $\,\{A\sigma A^*\}_{A\in\B}$ is a relatively compact subset of $\,\T_+(\K)$;}
  \item \emph{there exist an orthonormal basis $\{\varphi_i\}$ in $\H$, a non-degenerate probability distribution $\{p_i\}$ and a sequence $\{P_m\}$ of finite rank projectors in $\B(\K)$ such that}

   \begin{equation}\label{iii}
   \lim_{m\to+\infty}\sup_{A\in\B}\sum_i p_i \|(I_\K-P_m)A\varphi_i\|^2=0\quad \forall A\in\B.
   \end{equation}
\end{enumerate}

\emph{If equivalent properties $ (i)$-$(iii)$ are valid then}
\emph{\begin{itemize}
  \item $\{A\sigma A^*\}_{A\in\B}$ is a relatively compact subset of $\,\T_+(\K)$ for any state $\sigma\in\S(\H)$;
  \item relation (\ref{iii}) holds for arbitrary set $\{\varphi_i\}$ of unit vectors in $\H$ and any probability distribution $\{p_i\}$ provided that $\{P_m\}$ is
   an increasing sequence of finite rank projectors in $\B(\K)$ strongly converging to the unit operator $I_{\K}$.
\end{itemize}}
\end{property}\smallskip

\emph{Proof.} The implication $\rm (i)\Rightarrow(ii)$ is obvious. The implication $\rm (ii)\Rightarrow(i)$
follows from Corollary \ref{aqc-cr-c} in Section 2. Indeed, by using Lemma  \ref{1-rank} below it is easy to show
that the relative compactness of the family of quantum operations $\{A(\cdot)A^*\}_{A\in\B}$ in the strong convergence topology implies the relative compactness of $\B$ in the strong operator topology (we may assume that $\B$ lies within the unit ball of $\B(\H,\K)$).

The equivalence of $\rm(ii)$ and $\rm (iii)$ follows from the compactness criterion for bounded subsets of $\T_+(\K)$ \cite[Proposition 11]{AQC},
since any faithful state $\sigma$ in $\S(\H)$ has the representation $\,\sigma=\sum_ip_i|\varphi_i\rangle\langle\varphi_i|$, where
$\{\varphi_i\}$ is an orthonormal basis in $\H$ and $\{p_i\}$ is a non-degenerate probability distribution.

The first part of the last claim of the proposition is obvious, the second one can be easily proved by using Dini's lemma. $\Box$  \smallskip

\begin{lemma}\label{1-rank} \emph{If a sequence of quantum operations $\Phi_n(\cdot)=V_n(\cdot)V^*_n$ from $A$ to $B$ strongly converges
to a quantum operation $\Phi_0$ then there is a subsequence $\{V_{n_k}\}$ strongly converging to an operator $V_0:\H_A\to \H_B$ such that
$\,\Phi_0(\cdot)=V_0(\cdot)V^*_0$.}
\end{lemma}\smallskip

\emph{Proof.} By Remark \ref{note+} in Section 2 the operation $\Phi_0$ has the Choi rank $\leq 1$. So, $\Phi_0(\cdot)=U(\cdot)U^*$ for some contraction $U:\H_A\to \H_B$.

Since the unit ball of $\B(\H_A,\H_B)$ is compact in the weak operator topology we may assume (by passing to a subsequence)
that
\begin{equation}\label{w-o-c}
w.o.\shs\textup{-}\lim\limits_{n\rightarrow+\infty}V_n=V_0,
\end{equation}
where $V_0$ is a contraction in $\B(\H_A,\H_B)$.

By Lemma \ref{s-c-t-pr} in Section 3.1.4 the strong convergence of the sequence  $\{\Phi_n\}$ to the operation $\Phi_0$ implies
the strong convergence of the sequence  $\{\Phi_n\otimes \id_R\}$ to the operation $\Phi_0\otimes \id_R$, where $R$ is a quantum system
described by a separable Hilbert space $\H_R$. Note that (\ref{w-o-c}) implies that
\begin{equation}\label{w-o-c+}
w.o.\shs\textup{-}\lim\limits_{n\rightarrow\infty}V_n\otimes I_R=V_0\otimes I_R.
\end{equation}

Let $|\Omega\rangle=\sum_i \sqrt{p_i}\shs|\varphi_i\rangle\otimes |\psi_i\rangle$ be a unit vector in $\H_{AR}$ defined via a non-degenerate
probability distribution $\{p_i\}$ and given orthonormal base $\{\varphi_i\}$ and $\{\psi_i\}$ in $\H_A$ and $\H_R$, respectively.

By the strong convergence of the sequence $\{\Phi_n\otimes \id_R\}$ to the operation $\Phi_0\otimes \id_R$
the sequence $\{V_{n}\otimes I_R\,|\Omega\rangle\langle\Omega|\,V_{n}^*\otimes I_R\}_n$ tends to the operator $U\otimes I_R\,|\Omega\rangle\langle\Omega|\,U^*\otimes I_R$
in the trace norm. It follows that
\begin{equation}\label{t-rel+}
\lim_{n\to+\infty} \|V_{n}\otimes I_R|\Omega\rangle\|=\|U\otimes I_R|\Omega\rangle\|
\end{equation}
and there exists a sequence $\{\theta_n\}\subset[0,2\pi]$ such that
\begin{equation}\label{t-rel}
\lim_{n\to+\infty} e^{i\theta_n} V_{n}\otimes I_R|\Omega\rangle=U\otimes I_R|\Omega\rangle.
\end{equation}
Since the set $[0,2\pi]$ is compact there is a subsequence $\{\theta_{n_k}\}$ converging to $\theta_0\in[0,2\pi]$.
Using (\ref{t-rel}) it is easy to show that  the sequence $\{V_{n_k}\otimes I_R|\Omega\rangle\}_k$ converges
to the vector $e^{-i\theta_0}U\otimes I_R|\Omega\rangle$ in the norm of $\H_{BR}$.

At the same time, it follows from (\ref{w-o-c+}) that the sequence $\{V_{n_k}\otimes I_R\,|\Omega\rangle\}_k$ weakly converges
to the vector $V_0\otimes I_R|\Omega\rangle$ (as a sequence in the Hilbert space $\H_{BR}$). Thus, $V_0\otimes I_R|\Omega\rangle=e^{-i\theta_0}U\otimes I_R|\Omega\rangle$ and (\ref{t-rel+}) implies that
$\|V_{n_k}\otimes I_R|\Omega\rangle\|$ tends to $\|V_0\otimes I_R|\Omega\rangle\|$ as $k\to+\infty$. Hence, the sequence $\{V_{n_k}\otimes I_R\,|\Omega\rangle\}_k$  converges
to the vector $V_0\otimes I_R\,|\Omega\rangle$ in the norm of $\H_{BR}$ (by Theorem 1 in \cite[Section 26]{A&G}). Since
$$
V_{n_k}\otimes I_R\shs|\Omega\rangle=\sum_i \sqrt{p_i}\shs|V_{n_k}\varphi_i\rangle\otimes |\psi_i\rangle\quad \forall k\quad\textrm{and}\quad V_{0}\otimes I_R\shs|\Omega\rangle=\sum_i \sqrt{p_i}\shs|V_{0}\varphi_i\rangle\otimes |\psi_i\rangle,
$$
this implies that $V_{n_k}|\varphi_i\rangle$ tends to $V_{0}|\varphi_i\rangle$ as $k\to+\infty$ for all $i$. By noting that all the operators
$V_{n_k}$ and $V_{0}$ lie in the unit ball of $\B(\H_A,\H_B)$, we conclude that the sequence $\{V_{n_k}\}_k$ converges to the operator $V_0$ in the strong operator topology. $\Box$\medskip

Proposition \ref{new-com-cr}  provides an alternative way to prove many simple results concerning the strong convergence of sequences of operators in $\B(\H)$ (without using the standard arguments based on the notion of weak convergence in a Hilbert space). For example, to show that the strong convergence of a sequence $\{U_n\}$ of unitaries to a unitary operator $U_0$ implies the strong convergence of the sequence $\{U^*_n\}$ to the operator $U^*_0$ it suffices to note that the sequence $\{U_n^*U_0\sigma U_0^*U_n\}\subset\S(\H)$
tends to the state $U_0^*U_0\sigma U_0^*U_0=\sigma$ for any faithful state $\sigma$ in $\S(\H)$, since this proves, by Proposition \ref{new-com-cr}, the relative compactness of the sequence $\{U^*_n\}$ in the strong operator topology.\smallskip

Below we consider some non-trivial applications of Proposition \ref{new-com-cr}.\smallskip

\begin{example}\label{3-e}  Let $\{A_n\}$ be a sequence of operators from the  unit ball of the space $\B(\H,\K_1\otimes\K_2)$, where
$\H$, $\K_1$ and $\K_2$ are separable Hilbert spaces. By using Proposition \ref{new-com-cr} and Corollary 6 in \cite{AQC}
it is easy to show that \emph{the sequence $\{A_n\}$ is relatively compact in the strong operator topology if and only if
there is a faithful state $\sigma$ in $\S(\H)$ such that the sequences  $\{\Tr_{\K_2}A_n\sigma A_n^*\}\subset\T_+(\K_1)$ and
$\{\Tr_{\K_1}A_n\sigma A_n^*\}\subset\T_+(\K_2)$ are relatively compact.}

To show the usefulness of this condition assume that $\{\Phi_n\}$ is a sequence of quantum channels from $A$ to $B$ with bounded Choi rank
strongly converging to a channel $\Phi_0$. Let $\Phi_n(\rho)=\Tr_E V_n\rho V_n^*$ be the Stinespring representation of $\Phi_n$ for each $n\neq0$,
where $V_n:\H_A\to\H_{BE}$ is an isometry and $E$ is a finite-dimensional system. Let $\sigma$ be a faithful state in $\S(\H_A)$. As the sets
$\{\Tr_E V_n\sigma V_n^*\}_{n>0}$ and $\{\Tr_B V_n\sigma V_n^*\}_{n>0}$ are relatively compact (because of the strong convergence of $\Phi_n$ to $\Phi_0$ and due to the
compactness of $\S(\H_E)$, respectively), the above condition shows the relative compactness of the sequence $\{V_n\}$ in the strong operator topology.
So, there is an isometry $V_0:\H_A\to\H_{BE}$ and an increasing sequence $\{n_k\}$ of natural numbers such that
$$
s.o.\textup{-}\lim_{k\to+\infty}V_{n_k}=V_0\quad \textrm{and}\quad \Phi_0(\rho)=\Tr_E V_0\rho V_0^*,\quad \rho\in\S(\H_A).
$$
This means, roughly speaking, that \emph{from any sequence of Stinespring  representations of a strongly converging sequence
of quantum channels with bounded Choi rank it is possible to extract a subsequence converging to the Stinespring representation of a limit channel.}
This observation is dual and equivalent to the similar observation concerning the Kraus representations mentioned in Example \ref{l-2-e} in Section 3.1.

It is not hard to construct an example showing that the above claim is not valid if the Choi rank of the channels $\Phi_n$ is not uniformly bounded.
The Choi rank boundedness  condition can be replaced by the condition of relative compactness of the sequence of complementary channels $\,\widehat{\Phi}_n(\cdot)=\Tr_B V_n(\cdot)V_n^*$ in the strong convergence topology.
\end{example}\smallskip


\begin{example}\label{new-ex} Let  $\{P_m\}$ be
an increasing sequence of finite rank projectors in $\B(\H)$ strongly converging to the unit operator $I_{\H}$. For a given
vanishing sequence $\alpha_m$ consider the set
$$
\B_{P_m,\alpha_m}\doteq \{A\in\B(\H)\,|\,\|A\|\leq1, \|\bar{P}_mA-A\bar{P}_m\|\leq\alpha_m\},\quad \bar{P}_m=I_{\H}-P_m.
$$
The implication $\rm(iii)\Rightarrow (i)$ in Proposition \ref{new-com-cr} allows us to show that the set $\B_{P_m,\alpha_m}$ is compact in the strong convergence topology.
Indeed, it is easy to construct an orthonormal basis $\{\varphi_i\}$ in $\H$ such that $P_m=\sum_{i=1}^{\rank P_m}|\varphi_i\rangle\langle\varphi_i|$ for all $m$. Then for any probability
distribution $\{p_i\}$ and an arbitrary $A\in \B_{P_m,\alpha_m}$ we have
$$
\sum_{i=1}^{+\infty} p_i \|\bar{P}_mA\varphi_i\|^2\leq 2\sum_{i=1}^{+\infty} p_i\|A\bar{P}_m\varphi_i\|^2+2\sum_{i=1}^{+\infty} p_i\|\bar{P}_mA-A\bar{P}_m\|^2\leq
2\varepsilon_m+2\alpha^2_m,
$$
where $\varepsilon_m=\sum_{i=\rank P_m+1}^{+\infty} p_i=o(1)\,$ as $\,m\to+\infty$. So, the relation $\rm(iii)$ holds for the set $\B_{P_m,\alpha_m}$. The closedness
of this set in the strong convergence topology is obvious.\smallskip

Note that a direct proof of the above claim requires serious technical efforts.
\end{example}

\subsection{Petz's theorem for non-faithful states in infinite dimensions}

In this section we show how to use the compactness criterion from Corollary \ref{aqc-cr-c} to prove  that the famous
Petz theorem is valid for arbitrary states $\rho$ and $\sigma$ of infinite-dimensional quantum system including the case
when $\,\supp\rho\subsetneq\supp\sigma\,$ and there is no $c>0$ such that $c\rho\leq\sigma$. It seems that this case is not considered in the literature.\footnote{I would be grateful for any comments concerning this point.}
\smallskip

The \emph{quantum relative entropy} for two states $\rho$ and
$\sigma$ in $\mathfrak{S}(\mathcal{H})$ is defined as
\begin{equation}\label{qre-def}
D(\rho\shs\|\shs\sigma)=\sum_i\langle
\varphi_i|\,\rho\ln\rho-\rho\ln\sigma\,|\varphi_i\rangle,
\end{equation}
where $\{\varphi_i\}$ is the orthonormal basis of
eigenvectors of the state $\rho$ and it is assumed that
$D(\rho\|\shs\sigma)=+\infty$ if $\,\mathrm{supp}\rho\shs$ is not
contained in $\shs\mathrm{supp}\shs\sigma$ \cite{Ume,W,L-2}.\footnote{The support of a positive operator (in particular, state) is the orthogonal complement to its kernel.}\smallskip

Monotonicity of the quantum relative entropy means that
\begin{equation}\label{m-r-e}
    D(\Phi(\rho)\|\Phi(\sigma))\leq D(\rho\|\shs\sigma)
\end{equation}
for any quantum channel $\Phi:A\rightarrow B$ and any states
$\rho$ and $\sigma$ in $\S(\H_A)$.

Since the finiteness of $D(\rho\shs\|\shs\sigma)$ implies $\mathrm{supp}
\rho\subseteq\mathrm{supp} \sigma$, we will assume in what follows
that $\sigma$ and $\Phi(\sigma)$ are faithful states in $\S(\H_A)$
and in $\S(\H_B)$ correspondingly.

Petz's theorem characterizing the equality case in (\ref{m-r-e}) can
be formulated as follows.

\smallskip
\begin{theorem}\label{petz-t}
\emph{Let $D(\rho\shs\|\shs\sigma)<+\infty$. Then the equality holds in (\ref{m-r-e}) if and only if
$\,\Theta_{\sigma}(\Phi(\rho))=\rho$, where $\,\Theta_{\sigma}$ is the
channel from $\,B$ to $\,A$ defined by the formula}
\begin{equation}\label{theta}
\Theta_{\sigma}(\omega)=[\sigma]^{1/2}\Phi^*\left([\Phi(\sigma)]^{-1/2}\,\omega\,[\Phi(\sigma)]^{-1/2}\right)[\sigma]^{1/2},
\quad \omega\in\S(\H_B).
\end{equation}
\end{theorem}
\medskip
Note that $\Theta_{\sigma}(\Phi(\sigma))=\sigma$, so the above
criterion for the equality in (\ref{m-r-e}) can be treated as a
reversibility condition (sufficiency of the channel $\Phi$ with
respect to the states $\rho $ and $\sigma$ in terms of
\cite{P-sqc}).

Strictly speaking, the map $\Theta_{\sigma}$ is well defined by formula (\ref{theta}) on the set of states $\omega$ in $\S(\H_B)$,
for which  $[\Phi(\sigma)]^{-1/2}\,\omega\,[\Phi(\sigma)]^{-1/2}$ is a bounded operator. This always holds if the system $B$
is finite-dimensional, since we assume that $\Phi(\sigma)$ is a faithful state. The proof of (a generalized version of) Theorem \ref{petz-t} in the finite dimensional case can be found in \cite[the Theorem in Section 5.1]{H&Co}.

In infinite dimensions the finiteness of $D(\rho\|\shs\sigma)$ does not
imply that $c\rho\leq\sigma$ for some $c>0$ and hence
the argument of the map $\Phi^*$ in (\ref{theta}) with
$\omega=\Phi(\rho)$ may be an unbounded operator. Nevertheless, we
can define the channel $\Theta_{\sigma}$ as a predual map to the
linear completely positive normal unital map
\begin{equation}\label{theta-dual}
\Theta^*_{\sigma}(A)=[\Phi(\sigma)]^{-1/2}\Phi\left([\sigma]^{1/2}A\shs[\sigma]^{1/2}\right)[\Phi(\sigma)]^{-1/2},\quad
A\in\B(\H_A).
\end{equation}
This means that we can use formula (\ref{theta}), keeping in mind
that $\Phi^*$ is the extension of the map dual to $\Phi$ to unbounded operators
on $\H_B$ (which can be defined by $\Phi^*(\cdot)=\sum_k
V_k^*(\cdot)V_k$ via the Kraus representation $\Phi(\cdot)=\sum_k
V_k(\cdot)V_k^*$).\smallskip

With this definition of the channel $\Theta_{\sigma}$ Theorem
\ref{petz-t} is proved in \cite{P-sqc} (in the von Neumann
algebra settings and with the transition probability instead of the relative entropy)
under the condition that \emph{$\rho$ is a faithful state
in $\S(\H_A)$}.

If $\rho$ is a non-faithful state "dominated" by the state $\sigma$ in the sense that $c\rho\leq\sigma$ for some $c>0$
then the claim of Theorem \ref{petz-t} can be derived  from \cite[Theorem 2 and Proposition 4]{J&P}.\footnote{For details on the relationship of these results with Theorem \ref{petz-t}, see \cite[Ch.8,9]{O&P}.}
But, as noted above, this domination condition  does not hold for all  states $\rho$ such that $D(\rho\shs\|\shs\sigma)<\infty$.

To complete this gap, i.e. to prove that the claim of Theorem \ref{petz-t} is valid with
an \emph{arbitrary non-faithful} state $\rho$, one can use the approximating technique
based on the compactness criterion for families of quantum operations
in the strong convergence topology (presented in Corollary \ref{aqc-cr-c} in Section 2).

Consider the ensemble consisting of two states $\rho$ and $\sigma$
with probabilities $t$ and $1-t$, where $t\in(0,1)$. Let
$\sigma_{t}=t\rho+(1-t)\sigma$. By Donald's identity (\cite[Proposition
5.22]{O&P}) we have
\begin{equation}\label{d-one}
\!t D(\rho\|\,\sigma)=t D(\rho\|\,\sigma)+(1-t)D(\sigma\|\,\sigma)= t
D(\rho\|\,\sigma_{t})+(1-t)D(\sigma\|\,\sigma_{t})+D(\sigma_{t}\|\,\sigma)
\end{equation}
and
\begin{equation}\label{d-two}
\begin{array}{c}
t
D(\Phi(\rho)\|\Phi(\sigma))=t
D(\Phi(\rho)\|\Phi(\sigma))+(1-t)D(\Phi(\sigma)\|\Phi(\sigma))\\\\ \qquad\qquad=
 t
D(\Phi(\rho)\|\Phi(\sigma_{t}))+(1-t)D(\Phi(\sigma)\|\Phi(\sigma_{t}))+D(\Phi(\sigma_{t})\|\Phi(\sigma)).
\end{array}
\end{equation}
The left-hand sides of (\ref{d-one}) and (\ref{d-two}) are finite and coincide by the condition. So, since
the first, the second and the third terms in the right hand side of
(\ref{d-one}) are not less than the corresponding terms in
(\ref{d-two}) by monotonicity of the relative entropy, we conclude that
\begin{equation}\label{d-three}
D(\Phi(\rho)\|\Phi(\sigma_{t}))=D(\rho\|\sigma_{t})<+\infty\quad \forall t\in(0,1).
\end{equation}
Because the state $\rho$ is dominated by the state $\sigma_{t}$ for any $t\in(0,1)$  (as  $t\rho\leq\sigma_{t}$), it follows
from \cite[Theorem 2 and Proposition 4]{J&P} that (\ref{d-three}) implies that
$\rho=\Theta_{t}(\Phi(\rho))$ for all $t\in(0,1)$, where\footnote{Strictly speaking, $\Theta_{t}$ is the predual map to the
linear completely positive normal unital map $\Theta^*_{t}$ defined by the formula similar to (\ref{theta-dual}).}
$$
\Theta_{t}(\omega\,)=[\sigma_{t}]^{1/2}\Phi^*\left([\Phi(\sigma_{t})]^{-1/2}\shs\omega\shs[\Phi(\sigma_{t})]^{-1/2}\right)[\sigma_{t}]^{1/2}, \quad \omega\in\S(\H_B).
$$

To complete the proof it suffices to show that
\begin{equation}\label{s-lim-one}
\lim_{t\rightarrow +0}\Theta_{t}=\Theta_{\sigma}
\end{equation}
in the strong convergence topology, since this implies $\rho=\lim_{t\rightarrow
+0}\Theta_{t}(\Phi(\rho))=\Theta_{\sigma}(\Phi(\rho))$.

Since $\Theta_{t}(\Phi(\rho))=\rho$ and
$\Theta_{t}(\Phi(\sigma_t))=\sigma_t$ for all $t\in(0,1)$, we have $\Theta_{t}(\Phi(\sigma))=\sigma$ for all $t\in(0,1)$.
Thus, as $\Phi(\sigma)$ is a faithful state in $\S(\H_B)$, the set
$\{\Theta_{t}\}_{t\in(0,1)}$ of channels from $B$ to $A$ is relatively compact in
the strong convergence topology by Corollary \ref{aqc-cr-c} in Section 2. Hence,
there exists a sequence $\{t_n\}\subset(0,1)$ converging to zero such that
\begin{equation*}
\lim_{n\rightarrow+\infty}\Theta_{t_n}=\Theta_0
\end{equation*}
in the strong convergence topology, where $\Theta_0$ is a  channel form $B$ to $A$. By using simple arguments and the criterion (\ref{sc-ed}) of the strong convergence
one can show that $\Theta_0=\Theta_{\sigma}$ (see details in \cite[the Appendix]{HCD}). This proves (\ref{s-lim-one}).

\subsection{Preservation of reversibility under the strong convergence (direct proof) and beyond}

A quantum channel $\Phi:A\rightarrow B$ is called \emph{reversible}
with respect to a family $\S$ of states in $\S(\H_A)$ if there exists a quantum
channel $\,\Psi:B\rightarrow A$ such that
$\,\rho=\Psi\circ\Phi(\rho)\,$ for all $\,\rho\in\S$ \cite{J-rev,O&Co}. The channel $\Psi$ can be named  \emph{reversing} channel for $\Phi$.
This
property is also called sufficiency of the channel $\Phi$ for the family $\S$ \cite{P-sqc,J&P}.

By using  Petz's theorem (described in Section 3.3) and the lower  semicontinuity of the entropic disturbance as a function
of a pair (channel, input ensemble) one can show that \emph{the set of all quantum channels between  quantum systems $A$ and $B$
reversible w.r.t. a given family $\,\S$ of states in $\S(\H_A)$ is closed  w.r.t. the strong convergence} \cite[Corollary 17]{CSR}. It means
that for any sequence $\{\Phi_n\}$ of channels strongly converging to a channel $\Phi_0$ the following implication holds
$$
\forall n\;\; \exists \Psi_n: \rho=\Psi_n\circ\Phi_n(\rho)\;\;\forall\rho\in\S\quad \Rightarrow\quad \exists \Psi_0: \rho=\Psi_0\circ\Phi_0(\rho)\;\;\forall\rho\in\S.
$$
By using the compactness criterion  for
families of quantum channels in the strong convergence topology (presented in Corollary \ref{aqc-cr-c} in Section 2)  one can obtain a direct proof of this implication. Moreover, one can show that the reversing channel $\Psi_0$
can be always obtained as a limit point (in a certain sense) of the sequence $\{\Psi_n\}$.
\smallskip

In the following proposition we will denote the minimal subspace of $\H_B$ containing the supports of all the states $\Phi_0(\rho)$, $\rho\in\S(\H_A)$, by $\H_B^0$.
We will write $\Psi_n|_{\T(\H_B^0)}$ for the restriction of the map $\Psi_n:\T(\H_B)\to\T(\H_A)$ to the subspace $\T(\H_B^0)\subseteq\T(\H_B)$.

\smallskip
\begin{property}\label{r-ch} \emph{Let $\{\Phi_n\}$ be a sequence of channels from $A$  to $B$ reversible w.r.t. a family $\S\subseteq\S(\H_A)$.
Let $\{\Psi_n\}$ be the corresponding sequence of reversing channels, i.e. such channels  from $B$  to $A$ that
$\,\rho=\Psi_n\circ\Phi_n(\rho)\,$ for all $\,\rho\in\S$.}\smallskip

\emph{If the sequence $\{\Phi_n\}$ strongly converges to a channel $\Phi_0$ then
}\begin{itemize}
  \item \emph{the sequence $\{\Psi_n|_{\T(\H_B^0)}\}$ of channels from $\T(\H^0_B)$ to $\T(\H_A)$ is relatively compact in the strong convergence topology;}
  \item \emph{any partial limit $\Psi_*$ of the sequence $\{\Psi_n|_{\T(\H_B^0)}\}$ is a reversing channel for the channel $\Phi_0$ w.r.t. the family $\,\S$, i.e.
    $\,\rho=\Psi_*(\Phi_0(\rho))\,$ for all $\,\rho\in\S$.}
\end{itemize}
\end{property}

\textbf{Note:} The composition $\Psi_*\circ\Phi_0$ is well defined as the supports of all states at the output of $\Phi_0$ belong to the subspace $\H_B^0$ (by the definition of this subspace).\smallskip

\emph{Proof.} W.l.o.g. we may assume that the family $\S$ contains a faithful state $\rho_0$.
It is easy to show that $\supp\Phi_0(\rho_0)=\H_B^0$. Since
$$
\Psi_n(\Phi_n(\rho_0))=\rho_0\;\;\forall n\quad \textrm{and} \quad \lim_{n\to+\infty}\Phi_n(\rho_0)=\Phi_0(\rho_0),
$$
using the uniform boundedness of the operator norms of all the maps $\Psi_n$ it is easy to see that
$$
\lim_{n\to+\infty}\Psi_n(\Phi_0(\rho_0))=\rho_0.
$$
By Corollary \ref{aqc-cr-c} in Section 2
the above limit relation implies the relative compactness of the sequence of channels $\Psi^0_n\doteq\Psi_n|_{\T(\H_B^0)}$.
Let $\Psi_*$ be a partial limit of this sequence and $\{\Psi^0_{n_k}\}$ be its  subsequence strongly converging to $\Psi_*$.

Assume that $\sigma$ is an arbitrary state in $\S$ and denote the projector onto the subspace $\H_B^0$ by $P_0$. Since
$$
\lim_{k\to+\infty}P_0\Phi_{n_k}(\sigma)P_0=P_0\Phi_0(\sigma)P_0=\Phi_0(\sigma)=\lim_{k\to+\infty}\Phi_{n_k}(\sigma)
$$
and the operator norms of all the maps $\Psi_{n_k}$
are uniformly bounded, we have
$$
\lim_{k\to+\infty}\Psi_{n_k}(P_0\Phi_{n_k}(\sigma)P_0)=\Psi_{*}(P_0\Phi_0(\sigma)P_0)=\Psi_{*}(\Phi_0(\sigma))
$$
and
$$
\lim_{k\to+\infty}\Psi_{n_k}(P_0\Phi_{n_k}(\sigma)P_0)=\lim_{k\to+\infty}\Psi_{n_k}(\Phi_{n_k}(\sigma))=\sigma.
$$
The first limit relation follows from the strong convergence of the subsequence $\{\Psi^0_{n_k}\}$  to the channel $\Psi_*$, the second one is due to the
fact that  $\Psi_{n_k}(\Phi_{n_k}(\sigma))=\sigma$ for all $k$ because $\sigma\in\S$ and $\Psi_{n_k}$ is a reversing channel for $\Phi_{n_k}$.

These relations imply that  $\Psi_{*}(\Phi_0(\sigma))=\sigma$. So, $\Psi_*$ is a reversing channel for $\Phi_0$. $\Box$\smallskip

If the channel $\Phi_0$ in Proposition \ref{r-ch} is such that $\H_B^0=\H_B$ and $\Psi_0$ is a unique reversing channel for $\Phi_0$ then
Proposition \ref{r-ch} implies that the sequence $\{\Psi_n\}$ of reversing channels strongly converges to the channel $\Psi_0$.

\subsection{On existence of the Fawzi-Renner recovery channel reproducing the marginal
states in infinite dimensions}

\subsubsection{Preliminary facts}

The \emph{quantum conditional mutual information} (QCMI) of a state $\omega$ of a
finite-dimensional tripartite  quantum system $ABC$ is defined as
\begin{equation}\label{cmi-d}
    I(A\!:\!C|B)_{\omega}\doteq
    S(\omega_{AB})+S(\omega_{BC})-S(\omega)-S(\omega_{B}).
\end{equation}
This quantity plays important role in quantum
information theory \cite{H-SCI,Wilde}. The fundamental strong subadditivity property of the von Neumann
entropy means the nonnegativity of $\,I(A\!:\!C|B)_{\omega}$ \cite{Ruskai}.

The QCMI can be represented
by one of the formulae
\begin{equation}\label{cmi-d+}
    I(A\!:\!C|B)_{\omega}=I(A\!:\!BC)_{\omega}-I(A\!:\!B)_{\omega},
\end{equation}
\begin{equation}\label{cmi-d++}
    I(A\!:\!C|B)_{\omega}=I(AB\!:\!C)_{\omega}-I(B\!:\!C)_{\omega},
\end{equation}
where $I(X\!:\!Y)_{\omega}\doteq D(\omega_{XY}\|\omega_{X}\otimes\omega_{Y})$ is the mutual information
of the state $\omega_{XY}$ ($D(\cdot\|\cdot)$ is the quantum relative entropy defined in (\ref{qre-def})).
By these representations, the nonnegativity of $I(A\!:\!C|B)$ is a
direct corollary of the monotonicity of the relative entropy under
a partial trace.

If $\omega$ is a state of an infinite-dimensional tripartite quantum system $ABC$ then the right hand sides of (\ref{cmi-d}) and
of the representations (\ref{cmi-d+}) and (\ref{cmi-d++}) may contain the uncertainty $"\infty-\infty"$. In this case one can define the QCMI by one of the
following expressions
\begin{equation}\label{cmi-e+}
I(A\!:\!C|B)_{\omega}=\sup_{P_A}\left[\shs
I(A\!:\!BC)_{Q\omega Q}-I(A\!:\!B)_{Q\omega
Q}\shs\right]\!,\;\,Q=P_A\otimes I_B\otimes I_C,
\end{equation}
\begin{equation}\label{cmi-e++}
I(A\!:\!C|B)_{\omega}=\sup_{P_C}\left[\shs
I(AB\!:\!C)_{Q\omega Q}-I(B\!:\!C)_{Q\omega
Q}\shs\right]\!,\;\,Q=I_A\otimes I_B\otimes P_C,
\end{equation}
where the suprema are over all finite rank projectors
$P_X\in\B(\H_X),\, X\!=\!A,C$, and it is assumed that $I(X\!:\!Y)_{\sigma}=[\Tr
\sigma]I(X\!:\!Y)_{\sigma/\Tr\sigma}$ for any nonzero $\sigma$ in $\T_+(\H_{XY})$.

Expressions (\ref{cmi-e+}) and (\ref{cmi-e++}) are equivalent and coincide with the above formulae for any state $\omega$ at which these formulae are well defined. The QCMI defined by these  expressions is a nonnegative lower semicontinuous function on $\S(\H_{ABC})$ possessing all the basic properties of QMCI valid in the finite-dimensional case \cite[Theorem 2]{CMI}.

\subsubsection{The main result}

Fawzi and Renner proved in \cite{F&R}
that for any state $\omega$ of a tripartite quantum system $ABC$ there exists a recovery channel
$\Phi:B\rightarrow BC$ such that
\begin{equation}\label{FR-ineq}
2^{-\frac{1}{2}I(A:C|B)_{\omega}}\leq F(\omega,
\id_A\otimes\Phi(\omega_{AB})),
\end{equation}
where $F(\rho,\sigma)\doteq\|\sqrt{\rho}\sqrt{\sigma}\|_1$ is the
fidelity between states $\rho$ and
$\sigma$. This result can be considered as a
$\varepsilon$-version of the well-known characterization of a state
$\omega$ for which $I(A\!:\!C|B)_{\omega}=0$ as a Markov chain
(i.e. as a state reconstructed from its marginal state $\omega_{AB}$
by a channel $\id_A\otimes\Phi$). It has several important
applications in quantum information theory \cite{F&R,SFR}.\smallskip

It is also shown in Remark 5.3 in \cite{F&R} that in the finite-dimensional case  a channel $\Phi:B\rightarrow BC$ satisfying
(\ref{FR-ineq}) can be chosen in such a way that
\begin{equation}\label{FR-cond}
[\Phi(\omega_{B})]_B=\omega_{B}\quad\textrm{and}\quad
[\Phi(\omega_{B})]_C=\omega_{C},
\end{equation}
i.e. a recovery channel $\Phi$ may exactly reproduce the marginal
states.\smallskip

The existence of a channel $\Phi$ satisfying (\ref{FR-ineq}) is
proved in \cite{F&R} in the finite-dimensional settings by
quasi-explicit construction. Then, by using approximation technique,
this result is extended in \cite{F&R} (see also \cite{SFR}) to a state $\omega$
of infinite-dimensional system $ABC$ assuming that
$I(A\!:\!C|B)_{\omega}=S(A|B)_{\omega}-S(A|BC)_{\omega}$, i.e.
assuming that the marginal entropies of $\omega$ are finite. It is not hard
to update these arguments for arbitrary state $\omega$ of infinite-dimensional system $ABC$ assuming that
$I(A\!:\!C|B)_{\omega}$ is the extended QCMI defined by the equivalent expressions (\ref{cmi-e+}) and (\ref{cmi-e++}).\smallskip

The approximation technique based on the compactness criterion from Corollary \ref{aqc-cr-c} in Section 2 allows us to
extend the claim of Remark 5.3 in \cite{F&R} mentioned before to all states of
infinite-dimensional tripartite quantum systems.
\smallskip

\begin{property}\label{FR-r-m} \textbf{(ID-version of Remark 5.3 in
\cite{F&R})} \emph{For an arbitrary state $\omega$ of
an infinite-dimensional tripartite system $ABC$ there exists a channel
$\,\Phi:B\rightarrow BC$ satisfying (\ref{FR-ineq}) and
(\ref{FR-cond}) provided that $\,I(A\!:\!C|B)_{\omega}$ is the extended
quantum conditional mutual information (defined by the equivalent expressions (\ref{cmi-e+}) and (\ref{cmi-e++})).}
\end{property}\smallskip

The proof of this proposition (presented in \cite[Section 8.4]{CMI}) contains three basic steps in each of which
the relative compactness of some approximating sequence of quantum operations is established. The compactness criterion for
families of quantum operations in the strong convergence topology is used in this proof via the following\smallskip

\begin{lemma}\label{cmi-conv}
\emph{Let $\,\rho$ be a faithful state in $\,\S(\H_A)$ and
$\,\{\Phi_n\}$ be a sequence of quantum operations from $A$ to $BC$
such that
$$
[\Phi_n(\rho)]_B\leq \beta\quad \textit{and}\quad
[\Phi_n(\rho)]_C\leq \gamma\quad \forall n
$$
for some operators  $\,\beta\in\T_{+}(\H_B)$ and
$\,\gamma\in\T_{+}(\H_C)$. Then the sequence $\{\Phi_n\}$ is
relatively compact in the strong convergence topology.}
\end{lemma}\smallskip

\emph{Proof.} It suffices to note that the set
$\{\sigma\in\T_+(\H_{BC})\,|\,\sigma_B\leq\beta,
\sigma_C\leq\gamma\}$ is compact (by Corollary 6 in \cite[the Appendix]{AQC})
and to apply Corollary \ref{aqc-cr-c} in Section 2. $\square$
\medskip

\subsection{On closedness of the sets of degradable and anti-degradable channels w.r.t. the strong convergence}

There are two important classes of quantum channels defined via the notion of a complementary channel (described in Section 1).

A quantum channel $\Phi:A\to B$ is called \emph{degradable} if for any channel $\widehat{\Phi}:A\to E$ complementary to $\Phi$ there is
a channel $\Theta:B\to E$ such that $\widehat{\Phi}=\Theta\circ\Phi$ \cite{D-ch-0,D-ch}.

The well known property of  degradable
channels consists in the additivity of the coherent information, which implies that the quantum capacity of these channels is given by a single letter expression \cite{D-ch-0,H-SCI,Wilde-new}.
The private capacity of  degradable channels is also  given by a single letter expression and coincides with the quantum capacity \cite[Proposition 10.31]{H-SCI}. Another useful
property of  degradable channels is the lower semicontinuity, concavity and nonnegativity of the coherent information \cite{Wilde-new}.
\smallskip

A quantum channel $\Phi:A\to B$ is called \emph{anti-degradable} if for any channel $\widehat{\Phi}:A\to E$ complementary to $\Phi$ there is
a channel $\Theta:E\to B$ such that $\Phi=\Theta\circ\widehat{\Phi}$ \cite{D-ch}.
\smallskip

Since a complementary channel is defined up to the isometrical equivalence (see Section 1), to verify degradability (resp. anti-degradability) of a channel $\Phi$
it suffices to show that $\widehat{\Phi}=\Theta\circ\Phi$ (resp. $\Phi=\Theta\circ\widehat{\Phi}$)
for at least one channel $\widehat{\Phi}$ complementary to $\Phi$.\smallskip

\begin{property}\label{deg-ch} \emph{The sets $\,\F_{\rm d}(A,B)$ and $\,\F_{\rm a}(A,B)$  of degradable and anti-degradable
channels between arbitrary quantum systems $A$ and $B$ are closed w.r.t. the strong convergence.}
\end{property}\smallskip

\emph{Proof.} Let $\{\Phi_n\}$ be a sequence of channels in $\F_{\rm d}(A,B)$ strongly converging to a channel $\Phi_0$.
Let $\H_B^0$ be the minimal subspace of $\H_B$ containing the supports of all the states $\Phi_0(\rho)$, $\rho\in\S(\H_A)$.
If $\rho_0$ is any given faithful state in $\S(\H_A)$ then it is easy to show that $\H_B^0=\supp \Phi_0(\rho_0)$.

By Corollary 9A in \cite{CSR} there exist a system $E$ and a sequence $\{\Psi_n\}$ of channels from $A$ to $E$ strongly converging
to a channel $\Psi_0$ such that $\Psi_n=\widehat{\Phi}_n$ for all $n\geq0$.

Since $\Phi_n$ is a degradable channel for each $n>0$, there is a channel $\Theta_n:B\to E$ such that $\Psi_n=\Theta_n\circ\Phi_n$.
Because $\Phi_n(\rho_0)$ and $\Psi_n(\rho_0)=\Theta_n(\Phi_n(\rho_0))$ tend, respectively, to $\Phi_0(\rho_0)$ and $\Psi_0(\rho_0)$ as $n\to+\infty$, using the uniform boundedness of the operator norms of all the maps $\Theta_n$ it is easy to show that
$$
\lim_{n\to+\infty}\Theta_n(\Phi_0(\rho_0))=\Psi_0(\rho_0).
$$
Denote the restriction of the channel $\Theta_n$ to the subset $\T(\H_B^0)$ of $\T(\H_B)$ by $\Theta^0_n$. Write $B_0$ for a quantum system described by $\H_B^0$.
By Corollary \ref{aqc-cr-c} in Section 2
the above limit relation implies the relative compactness of the sequence $\{\Theta^0_n\}$ of channels from $B_0$ to  $E$. So, there exists a subsequence
$\{\Theta^0_{n_k}\}$ strongly converging to a channel $\Theta_*:B_0\to E$.

To prove that $\Phi_0$ is a degradable channel it suffices to show that $\Psi_0(\sigma)=\Theta_*(\Phi_0(\sigma))$ for any $\sigma\in\S(\H_A)$. We may apply the channel $\Theta_*$ to the state $\Phi_0(\sigma)$ as the support of this state belongs to the subspace $\H_B^0$ (by the definition of $\H_B^0$).

Denote the projector onto the subspace $\H_B^0$ by $P_0$. Since
$$
\lim_{k\to+\infty}P_0\Phi_{n_k}(\sigma)P_0=P_0\Phi_0(\sigma)P_0=\Phi_0(\sigma)=\lim_{k\to+\infty}\Phi_{n_k}(\sigma)
$$
and the operator norms of all the maps $\Theta_{n_k}$
are uniformly bounded, we have
$$
\lim_{k\to+\infty}\Theta_{n_k}(P_0\Phi_{n_k}(\sigma)P_0)=\Theta_{*}(P_0\Phi_0(\sigma)P_0)=\Theta_{*}(\Phi_0(\sigma))
$$
and
$$
\lim_{k\to+\infty}\Theta_{n_k}(P_0\Phi_{n_k}(\sigma)P_0)=\lim_{k\to+\infty}\Theta_{n_k}(\Phi_{n_k}(\sigma))=\lim_{k\to+\infty}\Psi_{n_k}(\sigma)=\Psi_{0}(\sigma),
$$
where the first (resp. the second)  limit relation follows from the strong convergence of the subsequence $\{\Theta^0_{n_k}\}$ (resp. $\{\Psi_{n_k}\}$) to the channel $\Theta_*$ (resp. $\Psi_0$).\smallskip

These relations imply that  $\Psi_{0}(\sigma)=\Theta_{*}(\Phi_0(\sigma))$. So, $\Phi_0$ is a degradable channel.\smallskip

Thus, the closedness of $\F_{\rm d}(A,B)$ is proved. To prove the closedness of $\F_{\rm a}(A,B)$ assume
that $\{\Phi_n\}$ is a sequence of channels in $\F_{\rm a}(A,B)$ strongly converging to a channel $\Phi_0$.
By Corollary 9A in \cite{CSR} there exist a system $E$ and a sequence $\{\Psi_n\}$ of channels from $A$ to $E$ strongly converging
to a channel $\Psi_0$ such that $\Psi_n=\widehat{\Phi}_n$ for all $n\geq0$. It follows that all the channels $\Psi_n$, $n>0$, are degradable.
By the above part of the proof $\Psi_0$ is a degradable channel. So, the channel $\Phi_0=\widehat{\Psi}_0$ is anti-degradable. $\Box$\smallskip

Proposition \ref{deg-ch} allows us to prove degradability (resp. anti-degradability) of a channel by representing this channel as a limit
of a strongly converging  sequence of degradable (resp. anti-degradable) channels.

\subsection{Preservation of convergence of the quantum relative entropy by quantum operations}

The following  theorem is proved in \cite{REC} by using the criterion of convergence (local continuity) of the quantum relative entropy
(obtained therein).\smallskip

\begin{theorem}\label{main}  \emph{Let $\,\{\rho_n\}$ and $\{\sigma_n\}$ be sequences of operators in $\,\T_+(\H_A)$ converging, respectively,
to operators  $\rho_0$ and $\sigma_0$ such that
\begin{equation*}
\lim_{n\to+\infty}D(\rho_n\|\shs\sigma_n)=D(\rho_0\|\shs\sigma_0)<+\infty.
\end{equation*}
Then
\begin{equation*}
\lim_{n\to+\infty}D(\Phi(\rho_n)\|\shs \Phi(\sigma_n))=D(\Phi(\rho_0)\|\shs \Phi(\sigma_0))<+\infty
\end{equation*}
for arbitrary quantum operation $\Phi:\T(\H_A)\to\T(\H_B)$.}
\end{theorem}\medskip

In this theorem $D(\varrho\shs\|\shs\varsigma)$ is  Lindblad's extension of the quantum relative entropy to any positive
operators $\varrho$ and
$\varsigma$ in $\mathfrak{T}(\mathcal{H})$ defined as
\begin{equation*}
D(\rho\shs\|\shs\sigma)=\sum_i\langle\varphi_i|\,\rho\ln\rho-\rho\ln\sigma+\sigma-\rho\,|\varphi_i\rangle,
\end{equation*}
where $\{\varphi_i\}$ is the orthonormal basis of
eigenvectors of the operator  $\varrho$ and it is assumed that $\,D(0\|\shs\varsigma)=\Tr\varsigma\,$ and
$\,D(\varrho\shs\|\varsigma)=+\infty\,$ if $\,\mathrm{supp}\varrho\shs$ is not
contained in $\shs\mathrm{supp}\shs\varsigma$ (in particular, if $\varrho\neq0$ and $\varsigma=0$)
\cite{L-2}. \smallskip

Theorem \ref{main} states that \emph{local continuity
of the quantum relative entropy is preserved by quantum operations}.  \smallskip

The compactness criterion for
families of quantum operations in the strong convergence topology (described in Section 2) along with the Stinespring representation of
strongly converging sequences of quantum channels obtained in \cite{CSR} allow us to strengthen the claim of Theorem \ref{main} as follows.  \smallskip

\begin{theorem}\label{main++} \emph{Let $\,\{\rho_n\}$ and  $\{\sigma_n\}$ be sequences of operators in $\,\T_+(\H_A)$ converging, respectively, to operators  $\rho_0$ and $\sigma_0$ such that}
\begin{equation*}
\lim_{n\to+\infty}D(\rho_n\|\shs\sigma_n)=D(\rho_0\|\shs\sigma_0)<+\infty.
\end{equation*}

\emph{If  $\,\{\varrho_n\}$ and $\{\varsigma_n\}$ are sequences of operators in $\,\T_+(\H_B)$ converging, respectively, to operators  $\varrho_0$ and $\varsigma_0$
such that $\varrho_n=\Phi_n(\rho_n)$ and $\varsigma_n=\Phi_n(\sigma_n)$ for each $n\neq0$, where $\Phi_n$ is a quantum operation from $A$ to $B$,  then}
\begin{equation*}
\lim_{n\to+\infty}D(\varrho_n\|\shs\varsigma_n)=D(\varrho_0\|\shs\varsigma_0)<+\infty.
\end{equation*}
\end{theorem}\smallskip

It is essential that in Theorem \ref{main++} no properties of the sequence $\{\Phi_n\}$ are assumed.
The proof of Theorem \ref{main++} presented in \cite[Section 5.2]{REC} is based on using Lemma \ref{l-1} in Section 3.1.

\section*{The Appendix: The compactness criterion for subsets of positive linear maps between
spaces of trace-class operators in the strong convergence topology}

Let $\mathfrak{L}_{+}(A,B)$ be the cone of positive
linear bounded maps from  the Banach space
$\mathfrak{T}(\mathcal{H}_A)$ of trace-class operators on a separable
Hilbert space $\mathcal{H}_A$ into the analogous Banach space
$\mathfrak{T}(\mathcal{H}_B)$. The \emph{strong convergence topology} on $\mathfrak{L}_{+}(A,B)$ is defined by the family of seminorms $\Phi\mapsto\|\Phi(\rho)\|_1$, $\rho\in\S(\H_A)$. The convergence of a sequence $\{\Phi_n\}$ of maps $\mathfrak{L}_{+}(A,B)$  to a map $\Phi_0\in\mathfrak{L}_{+}(A,B)$ in this topology means
the validity of the limit relation (\ref{star+}) for any state $\rho$ in $\S(\H_A)$. It is clear that
the topology of strong convergence on $\mathfrak{L}_{+}(A,B)$ is the restriction to $\mathfrak{L}_{+}(A,B)$ of the strong operator topology on the space
of all bounded linear maps between $\mathfrak{T}(\mathcal{H}_A)$ and $\mathfrak{T}(\mathcal{H}_B)$.\smallskip

\begin{property}\label{comp-c} 
\textup{A)} \textit{A closed bounded subset
$\,\mathfrak{L}_{0}\subseteq\mathfrak{L}_{+}(A,B)$
is compact in the strong convergence topology if there exists a  faithful state
$\sigma$ in
$\mathfrak{S}(\mathcal{H}_A)$ such that $\{\Phi(\sigma)\}_{\Phi\in\mathfrak{L}_{0}}$ is a
compact subset of $\mathfrak{T}(\mathcal{H}_B)$. }\smallskip

\textup{B)} \textit{If a subset
$\mathfrak{L}_{0}\subseteq\mathfrak{L}_{+}(A,B)$
is compact in the strong convergence topology then
$\{\Phi(\sigma)\}_{\Phi\in\mathfrak{L}_{0}}$ is a compact subset of
$\mathfrak{T}(\mathcal{H}_B)$ for any state $\sigma$ in
$\mathfrak{S}(\mathcal{H}_A)$.}
\end{property}
\smallskip

\emph{Proof.} A) Let $\{|i\rangle\}$ be the basis of
eigenvectors of the state $\sigma$ corresponding to the sequence of its eigenvalues arranged in the non-increasing order
and $\mathcal{H}_{m}$ be the subspace generated by the first
$m$ vectors of this basis.\smallskip

Let $\{\Phi_{n}\}$ be an arbitrary sequence of maps in
$\mathfrak{L}_{0}$.\smallskip

Show that for each natural $m$ and arbitrary operator $\rho$ in
$\mathfrak{T}(\mathcal{H}_{m})$ there exists a  subsequence
$\{\Phi_{n_{k}}\}$ such that the sequence $\{\Phi_{n_{k}}(\rho)\}_{k}$
has a limit in $\mathfrak{T}(\mathcal{H}_B)$. Suppose first that $\rho\geq
0$. Since $\rho\in\mathfrak{T}(\mathcal{H}_{m})$ there exists such
$c_{\rho}>0$ that $c_{\rho}\rho\leq\sigma$. By the compactness
criterion for subsets of $\mathfrak{T}(\H_B)$ (Proposition 11 in \cite[the Appendix]{AQC}) for arbitrary
$\varepsilon>0$ there exists a finite rank (orthogonal) projector
$P_{\varepsilon}\in\mathfrak{B}(\H_B)$ such that
$\mathrm{Tr}(I_B-P_{\varepsilon})\Phi(\sigma)<\varepsilon$,
and hence
$\mathrm{Tr}(I_B-P_{\varepsilon})\Phi(\rho)<c_{\rho}^{-1}\varepsilon$
for all $\Phi\in\mathfrak{L}_{0}$. So, by the same compactness criterion
the sequence $\{\Phi_n(\rho)\}$ is relatively compact. This
implies the existence of a subsequence with the required properties for any positive operator
$\rho$. The existence of such subsequence for an arbitrary operator
$\rho\in\mathfrak{T}(\mathcal{H}_{m})$ follows from the representation of
this operator as a linear combination of four positive operators in
$\mathfrak{T}(\mathcal{H}_{m})$.

Thus, for each natural $m$ an arbitrary sequence
$\{\Phi_{n}\}\subset\mathfrak{L}_{0}$ contains a subsequence
$\{\Phi_{n_{k}}\}$ such that
\begin{equation}\label{lim-exp}
\exists\lim_{k\rightarrow+\infty}\Phi_{n_{k}}(|i\rangle\langle
j|)=\omega^{m}_{ij}
\end{equation}
for all $i,j=\overline{1,m}$, where $\{\omega^{m}_{ij}\}$ are
operators in $\mathfrak{T}(\mathcal{H}_B)$.

For arbitrary  $m'>m$, by applying the above observation to the
sequence $\{\Phi_{n_{k}}\}_{k}$, we obtain a subsequence of the
sequence $\{\Phi_{n}\}$ such that (\ref{lim-exp}) holds for all
$i,j=\overline{1,m'}$ with a set of operators  $\{\omega^{m'}_{ij}\}$
such that $\omega^{m'}_{ij}=\omega^{m}_{ij}$ for all $i,j=\overline{1,m}$.

By repeating  this construction one can show the existence of the set
$\{\omega_{ij}\}_{i,j=1}^{+\infty}$ of operators in $\T(\H_B)$ having the following
property: for each $m$ there exists a subsequence
$\{\Phi_{n_{k}}\}$ of the sequence $\{\Phi_{n}\}$ such that
(\ref{lim-exp}) holds with $\omega^{m}_{ij}=\omega_{ij}$ for all
$i,j=\overline{1,m}$.

Consider the map on the set
$\bigcup_{m\in\mathbb{N}}\mathfrak{T}(\mathcal{H}_{m})$ defined as
follows
$$
\Phi_{*}\, :\, \sum_{i,j} a_{ij}\, |i\rangle\langle j|\quad \mapsto
\quad \sum_{i,j} a_{ij}\, \omega_{ij}\ \in\ \mathfrak{T}\,
(\mathcal{H}_B).
$$
This map is linear by construction. It is easy to see its positivity
and boundedness. Indeed, by the property of the set $\{\omega_{ij}\}$ for
arbitrary operator $\rho\in\bigcup_{m}\mathfrak{T}(\mathcal{H}_{m})$
there exists a subsequence $\{\Phi_{n_{k}}\}$ of the sequence
$\{\Phi_{n}\}$ such that $\, \Phi_{*}(\rho)\, =\,
\lim\limits_{k\rightarrow+\infty}\Phi_{n_{k}}(\rho)$. Thus, the positivity
and boundedness of the map $\Phi_{*}$ follow from the positivity of the
maps in the sequence $\{\Phi_{n}\}$ and their uniform boundedness.
Since the set $\bigcup_{m}\mathfrak{T}(\mathcal{H}_{m})$ is dense in
$\mathfrak{T}(\mathcal{H}_A)$, the map $\Phi_{*}$ can be extended to
a linear positive bounded map from $\mathfrak{T}(\mathcal{H}_A)$ into
$\mathfrak{T}(\H_B)$ (denoted by the same symbol
$\Phi_{*}$).

Show that the map $\Phi_{*}$ is a limit point of the sequence
$\{\Phi_{n}\}$ in the strong convergence topology. This topology on
bounded subsets of $\mathfrak{L}_{+}(A,B)$ can
be determined by a countable family
$\Phi\mapsto\|\Phi(\rho)\|_{1}$, $\rho\in\S_0$, of seminorms, where
$\S_0$ is any countable dense subset of the set
$\mathfrak{S}(\mathcal{H}_A)$.\footnote{Here the possibility to
express arbitrary operator in $\mathfrak{T}(\mathcal{H}_A)$ as linear
combination of four states in $\mathfrak{S}(\mathcal{H}_A)$ is used.} It is clear that we may  choose  the  subset $\S_0$ consisting of states in
$\bigcup_{m}\mathfrak{T}(\mathcal{H}_{m})$. An arbitrary vicinity of
the map $\Phi_{*}$ contains  vicinity of the form
$$
\left\{\left.\Phi\in\mathfrak{L}_+(A,B)\,\right|\,\|(\Phi-\Phi_{*})(\rho_{i})\|_{1}<\varepsilon,
i=\overline{1,p}\shs\right\},\quad p\in\N,
$$
where $\{\rho_{i}\}_{i=1}^{p}$ is a finite subset of $\S_0$ and $\varepsilon>0$. Since
$\{\rho_{i}\}_{i=1}^{p}\subset\mathfrak{T}(\mathcal{H}_{m})$ for
a particular $m$, the construction of the map $\Phi_{*}$ implies
the existence of a subsequence $\{\Phi_{n_{k}}\}$ of the sequence
$\{\Phi_{n}\}$ such that
$\Phi_{*}(\rho_{i})=\lim_{k\rightarrow+\infty}\Phi_{n_{k}}(\rho_{i})$
for all $i=\overline{1,p}$. This shows the existence of at least one
element of the sequence $\{\Phi_{n}\}$ in the above vicinity.

Thus, the map $\Phi_{*}$ is a limit point of the sequence
$\{\Phi_{n}\}$ in the strong convergence topology. By metrizability of
the strong convergence topology on bounded subsets of the cone
$\mathfrak{L}_{+}(A,B)$ this implies the existence
of a subsequence of the sequence $\{\Phi_{n}\}$ strongly converging to the
map $\Phi_{*}$.\footnote{Another way to prove this is to use the "diagonal" method right after the definition of the map $\Phi_*$.} Compactness of the set $\mathfrak{L}_{0}$ is proved.\smallskip

B) Since the compactness is preserved under action of continuous maps, this assertion obviously follows from the definition of the
strong convergence topology. $\square$

\bigskip
I am grateful to A.S.Holevo and V.Zh.Sakbaev for useful discussion. Special thanks to M.M.Wilde for  valuable communication.


\bigskip

\begin{thebibliography}{99}

\bibitem{Kit} D.Aharonov, A.Kitaev, N.Nisan, "Quantum circuits with mixed states", Proc. 30th STOC,
pp. 20-30, ACM Press, 1998.

\bibitem{A&G}  N.I.Akhiezer, I.M.Glazman, "Theory of Linear Operators in Hilbert Space", Dover Publications (1993).

\bibitem{B&R} O.Bratteli, D.W.Robinson, "Operators algebras and quantum statistical mechanics", vol.I, Springer Verlag, New York-Heidelberg-Berlin, 1979.

\bibitem{Choi} M.D.Choi,  "Completely positive linear maps on complex matrices", Linear Algebra and Its
Application, 10, 285-290 (1975).

\bibitem{D-ch} T.S.Cubitt, M.B.Ruskai, G.Smith,  "The structure of degradable quantum channels", J. Math. Phys. 49, 102-104 (2008).

\bibitem{D-A} G.F.Dell'Antonio, "On the limits of sequences of normal states", Commun.
Pure Appl. Math. 20, 413-430, 1967.

\bibitem{D-ch-0} I.Devetak, P.W.Shor, "The capacity of a quantum channel for simultaneous transmission of classical and quantum information",
Commun. Math. Phys. 256:2, 287-303 (2005); arXiv:quant-ph/0311131.


\bibitem{F&R} O.Fawzi, R.Renner, "Quantum conditional mutual information and
approximate Markov chains", Comm. Math. Phys., 340:2 (2015), 575-611; arXiv:1410.0664.

\bibitem{H&Co} F.Hiai, M.Mosonyi, D.Petz, C.Beny "Quantum f-divergences and error correction", Reviews in Mathematical Physics  23:7, 691-747 (2011);
arXiv:1008.2529.

\bibitem{H-SCI} A.S.Holevo "Quantum systems, channels, information.
A mathematical introduction", Berlin, DeGruyter, 2012.

\bibitem{H-SSQT} A.S.Holevo, "Statistical structure of quantum theory", Lect. Notes Phys., 67, Springer, 2001.

\bibitem{H&W} A.S.Holevo, R.F.Werner, "Evaluating capacities of Bosonic Gaussian channels", Phys. Rev. A, 63 (2001), 032312.

\bibitem{H-c-w-c} A.S.Holevo "Classical capacities of quantum channels with
constrained inputs", Probability Theory and Applications, 48:2,
359-374, (2003); quant-ph/0211170;

\bibitem{H-c-ch} A.S.Holevo, "Complementary channels and the additivity problem", Theory Probab. Appl., 51:1, 92-100 (2007).

\bibitem{Jam} A.Jamiolkowski, "Linear transformations which preserve trace and positive semidefiniteness of operators",
Rep. Math. Phys. 3, pp. 275-278 (1972)

\bibitem{J&P} A.Jencova, D.Petz "Sufficiency in quantum statistical inference",
Commun. Math. Phys. 263, 259-276 (2006); arXiv:quant-ph/0604091.

\bibitem{J-rev} A.Jencova, "Reversibility conditions for quantum operations",  Rev. Math. Phys. 24,  1250016 (2012).

\bibitem{Wilde-new} S.Khatri, M.M.Wilde, "Principles of Quantum Communication Theory: A Modern Approach", arXiv:2011.04672.

\bibitem{L&Co} L.Lami, B.Regula, "Computable lower bounds on the entanglement cost of quantum channels", J. Phys. A: Math. Theor. 56, 035302 (2023); arXiv:2201.09257.

\bibitem{Ruskai} E.H.Lieb, M.B.Ruskai, "Proof of the strong suadditivity of quantum
mechanical entropy", J.Math.Phys. 14 1938 (1973).

\bibitem{L-2} G.Lindblad  "Expectation and Entropy Inequalities for Finite
Quantum Systems", Comm. Math. Phys. 39:2, 111-119 (1974).

\bibitem{O&Co} T.Ogawa, A.Sasaki, M.Iwamoto, H.Yamamoto, "Quantum Secret Sharing Schemes and Reversibility of Quantum Operations", Phys. Rev. A 72, 032318 (2005).

\bibitem{O&P} M.Ohya, D.Petz, Quantum Entropy and Its Use, Theoretical and Mathematical Physics (Springer Berlin Heidelberg,
2004).

\bibitem{P-sqc} D. Petz "Sufficiency of channels over von Neumann algebras", Quart. J. Math. Oxford
Ser. (2) 39:153, 97-108 (1988).

\bibitem{Pir} S.Pirandola, R.Laurenza, S.L.Braunstein, "Teleportation simulation of bosonic Gaussian channels: Strong and uniform convergence",
Eur. Phys. J. D 72, 162, (2018).

\bibitem{P&Sh} V.Yu.Protasov, M.E.Shirokov, "Generalized compactness in linear spaces and its applications", Sb. Math., 200:5 (2009), 697-722; arXiv: 1002.3610.

\bibitem{R&S} M.Reed, B.Simon, "Methods of Modern Mathematical Physics. Vol I. Functional Analysis", Academic Press Inc., 1980.

\bibitem{SFR} D.Sutter, O.Fawzi, R.Renner
"Universal recovery map for approximate Markov chains", Proc. A, 472:2186 (2016), 20150623, 26 pp.;
arXiv:1504.07251.

\bibitem{BSimon} B.Simon, "Operator Theory: A Comprehensive Course in Analysis, Part IV",  American Mathematical Society, 2015.

\bibitem{AQC} M.E.Shirokov, A.S.Holevo  "On approximation of
infinite dimensional quantum channels", Problems of Information
Transmission. 44:2, 3-22 (2008); arXiv: 0711.2245.

\bibitem{HCD} M.E.Shirokov, "Monotonicity of the Holevo quantity: a necessary condition for equality in terms of a channel and its applications", arXiv:1106.3297.

\bibitem{CMI} M.E.Shirokov, "Measures of correlations in infinite-dimensional quantum systems", Sb. Math., 207:5 (2016), 724-768; arXiv:1506.06377.

\bibitem{REC} M.E. Shirokov, "Convergence criterion for quantum relative entropy and its use", Sb. Math., 213:12 (2022), 1740-1772  (PDF file available at
 https://doi.org/10.4213/sm9794e)

\bibitem{CSR} M.E.Shirokov, "Strong convergence of quantum channels: continuity of the Stinespring dilation and discontinuity of the unitary dilation", J. Math. Phys. 61, 082204 (2020).

\bibitem{Ume} H.Umegaki, "Conditional expectation in an operator algebra, IV (entropy and information)", Kodai Math.Sem.Rep., 14, 59-85 (1962).

\bibitem{W} A.Wehrl, "General properties of entropy", Rev. Mod. Phys. 50,
221-250 (1978).

\bibitem{Wilde} M.M.Wilde, "Quantum Information Theory".  Cambridge University Press, 2013.

\bibitem{Wilde-sc} M.M.Wilde, "Strong and uniform convergence in the teleportation simulation of bosonic Gaussian channels", Physical Review A, 97:6, 062305 (2018).

\bibitem{W-EBN}  A.Winter, "Energy-Constrained Diamond Norm with Applications to the Uniform Continuity of Continuous
Variable Channel Capacities", arXiv:1712.10267 (2017).














\end{thebibliography}
\end{document}